\documentclass[aps,prl,superscriptaddress]{revtex4-2} %,twocolumn

\usepackage{amsmath,amssymb}
\usepackage[T1]{fontenc}
\usepackage{graphicx}% Include figure files
\usepackage{dcolumn}% Align table columns on decimal point
\usepackage{bm}% bold math
\usepackage{bbm}
\usepackage{siunitx} % (SI) units package
\usepackage{xcolor}
\usepackage{listings}
\usepackage{lineno}

\usepackage{pdfpages}
\usepackage{pgffor}
\makeatletter\AtBeginDocument{\let\LS@rot\@undefined}\makeatother

\begin{document}
%\linenumbers

\title{Generative modelling powered by room-temperature polariton condensates}

\author{Yuan Wang}
\affiliation{School of Mathematical and Physical Sciences, University of Sheffield, Sheffield S10 2TN, United Kingdom}

\author{Marcin Muszynski}
\affiliation{Department of Physics, City College of New York, New York, NY 10031, USA}

\author{Avinash Dash}
\affiliation{Department of Physics, City College of New York, New York, NY 10031, USA}
\affiliation{Physics Doctoral Program, Graduate Center of the City University of New York, New York, NY 10016, USA}

\author{Rishabh Kaurav}
\affiliation{Department of Physics, City College of New York, New York, NY 10031, USA}
\affiliation{Chemistry Doctoral Program, Graduate Center of the City University of New York, New York, NY 10016, USA}

\author{Vinod M. Menon}
\email{vmenon@ccny.cuny.edu}
\affiliation{Department of Physics, City College of New York, New York, NY 10031, USA}
\affiliation{Physics Doctoral Program, Graduate Center of the City University of New York, New York, NY 10016, USA}

\author{Oleksandr Kyriienko}
\email{o.kyriienko@sheffield.ac.uk}
\affiliation{School of Mathematical and Physical Sciences, University of Sheffield, Sheffield S10 2TN, United Kingdom}

\begin{abstract}
Generative modelling requires efficient stochastic nonlinear transformations and physical platforms that can naturally realise them. We experimentally demonstrate that nonlinear optical systems operating in the strong light-matter coupling regime can serve as physical transformation layers for conditional generative modelling. Specifically, we develop a workflow in which room-temperature exciton-polariton condensates formed in organic dye microcavities act as a physical stochastic transform within a generative adversarial network and enable conditional digit-to-image translation. By using the nonlinear many-body dynamics and intrinsic stochasticity of polariton condensates, the workflow outperforms baseline approaches based on digitally injected perturbations. We find that polariton-enabled sampling via generative adversarial network (Polariton GAN) yields improved inception score, digit preservation accuracy and structural similarity compared with both digital sampling and laser-based systems. We further show that spatially correlated output variations can naturally regularise adversarial training and enhance output diversity. Our results establish polariton condensation as a new computational resource for generative modelling, opening a pathway towards physics-enhanced machine learning systems.
\end{abstract}

\maketitle

\section{Introduction}

Modern generative models rely on learning stochastic nonlinear maps from simple, structured inputs to complex data distributions. This capability underpins image synthesis, language generation and inverse design in chemistry and materials science~\cite{goodfellow2014generative,ho2020denoising,Vaswani2017attention,brown2020language,sanchez2018inverse,yao2021inverse,wang2023scientific}. But the same progress has also exposed a growing hardware problem: state-of-the-art AI is increasingly constrained by the energy and infrastructure cost of scale~\cite{IEA2025}. This has intensified interest in physical machine-learning platforms that do not merely accelerate digital workloads, but contribute useful transformations directly through their underlying physical operation~\cite{tanaka2019recent,shastri2021photonics,mcmahon2023physics,aifer2025solving}.

Photonics is a natural candidate for this role because it combines high bandwidth, low latency and massive parallelism~\cite{shastri2021photonics,mcmahon2023physics}. These strengths have already enabled optical machine-learning hardware for image sensing, neural inference and large-scale linear transformations~\cite{wetzstein2020inference,shen2017deep,harris2018linear,feldmann2021parallel,wang2023image,hua2025integrated,ma2026machine}. More recently, photonic platforms have started to move towards probabilistic and generative settings, including noise-enabled and sampling-based schemes~\cite{choi2024photonic,bruckerhoff2024probabilistic,wu2022harnessing,zhan2024photonic,chen2025optical}. Yet a central bottleneck remains: photonic hardware is especially mature for matrix-vector multiplication, interference and linear projection, whereas modern generative modelling depends crucially on nonlinear transformations together with a controllable source of variability~\cite{wang2023image,yanagimoto2026programmable}. The question is therefore not only how to accelerate generative models optically, but which part of the generative workflow can be delegated most naturally to an optical system.

Exciton polaritons are appealing in this context because they offer exactly this combination of optical accessibility, nonlinearity and stochastic driven-dissipative dynamics. Arising from strong coupling between cavity photons and excitons~\cite{deng2010exciton,basov2025polaritonic}, they support condensation, collective nonlinear behaviour and quantum-fluid-like dynamics in space and time~\cite{carusotto2013quantum,kasprzak2006bose,keeling2020bose,amo2009collective,sich2012observation}. Strong interactions and nonlinear response have been explored across several polaritonic material platforms~\cite{togan2018enhanced,munoz2019emergence,delteil2019towards,kuriakose2022few,yagafarov2020mechanisms,kyriienko2020nonlinear,song2024microscopic,struve2026room}. Among them, organic microcavities are particularly attractive because they support room-temperature operation through the large binding energies and strong oscillator strengths of Frenkel excitons~\cite{kena2010room,plumhof2014room,zhao2022nonlinear,deshmukh2024plug,georgakilas2025room}. They can host polariton lasing and condensation in engineered spatial patterns, while exhibiting pronounced nonlinear response associated with saturation and phase-space filling~\cite{dusel2020room,betzold2024dirac,yagafarov2020mechanisms}.

To date, polariton-based machine learning was developed mainly in regimes where the physical system acts as a classifier, reservoir or feature generator~\cite{opala2023harnessing,ghosh2019quantum,ghosh2020reconstructing,ghosh2021quantum}. Exciton-polariton lattices have been used for neuromorphic image classification~\cite{ballarini2020polaritonic,mirek2021neuromorphic}. Room temperature polaritonic platforms were developed for reservoir computing and digit recognition~\cite{kkedziora2024predesigned,opala2024room,zaremba2025optically,gan2025ultrafast}. Recent hybrid workflows have also used polaritonic dynamics to produce graph-aware embeddings for qualitatively improved learning~\cite{wang2025polaritonic,wang2025photonics}. Collectively, this suggests that polaritonic systems are well suited to supplying nonlinear physical preprocessing within a broader digital pipeline. What remains largely unexplored is whether their intrinsic stochasticity and spatially correlated nonlinear response can be used directly as a computational resource for generative modelling, rather than treated as noise to be averaged away.

Here, we demonstrate such a workflow using room-temperature polariton condensates as a physical stochastic nonlinear layer within a conditional generative adversarial network (GAN). Our task is digit-to-image translation: simple input digit patterns are mapped to handwritten-style outputs while preserving digit identity. The key idea is to place the polaritonic system at the stochastic transformation stage of the generative pipeline. Instead of injecting variability digitally as an abstract random perturbation, we use the nonlinear many-body dynamics and shot-to-shot fluctuations of an exciton-polariton condensate to generate the physically transformed inputs. 
We show that this polariton-assisted workflow improves generative performance relative to digital perturbation baseline, and is distinct from laser-based control experiment. In particular, polariton-processed inputs lead to more stable adversarial training, stronger digit preservation, and greater output diversity. The results identify polariton condensation as a practical room-temperature photonic resource for generative modelling and suggest a broader route towards physics-enhanced machine learning systems in which nonlinear optical hardware contributes directly to sampling and stochastic transformation.
%%% 
\begin{figure*}[h]
\begin{center}
\includegraphics[scale=1]{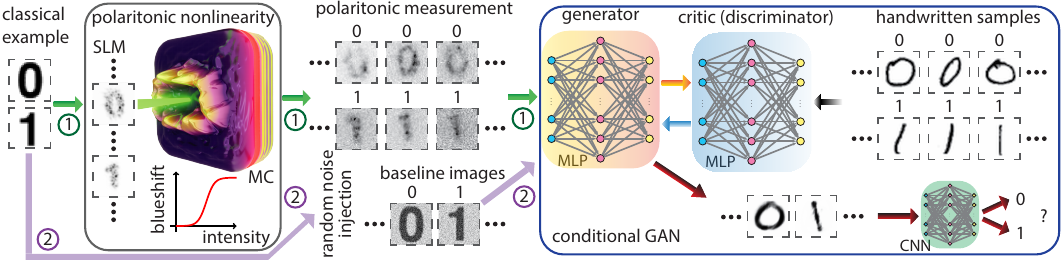}
\end{center}
\caption{\textbf{Workflow for polaritonic conditional generative adversarial network (GAN) for digit-to-image translation.} The system operates in two training modes. In the condensate mode (\raisebox{.5pt}{\textcircled{\raisebox{-.9pt} {1}}}, green arrows) classical digit examples (`0' or `1') are encoded using a spatial light modulator (SLM) and injected into a polariton condensate system. The resulting photoluminescence (PL) produces transformed digit images through nonlinear condensate dynamics. In the baseline mode (\raisebox{.5pt}{\textcircled{\raisebox{-.9pt} {2}}}, purple arrows), random noise is used without any physical processing. Both modes together with their corresponding labels (shown on the top of each image) feed into the conditional GAN framework, where a multilayer perceptron (MLP) generator learns to produce MNIST (handwritten)-style digit images conditioned on the input digit label (`0' or `1'). The MLP critic (discriminator) evaluates generated samples against real MNIST handwritten digits, providing adversarial feedback that drives the generator to produce more realistic outputs. During training, the critic updates its parameters to better distinguish real from generated samples, while the generator adapts to fool the critic by improving output quality. This minimax game continues iteratively until convergence. The trained system produces digit patterns that are subsequently validated by a pre-trained convolutional neural network (CNN) classifier to ensure digit identity preservation (distinguishing between `0' and `1').}
\label{fig:fig1}
\end{figure*}
%%%

\section{Model}\label{sec-mod}

We consider a conditional image-generation task in which an input digit pattern is mapped to a handwritten-style output while preserving its label, also referred to here as digit-to-image translation. To assess the role of nonlinear polariton dynamics as a stochastic transformation, we compare two input modes: a condensate-based approach, in which the input is transformed by polariton dynamics, and a digital baseline based on adding random Gaussian perturbations to standard Helvetica digits (see Methods for details). In addition to the condensate and digital baseline modes, we also consider a laser mode as a control, in which below-threshold optical patterns with similar spatial structure are combined with computational perturbations, allowing us to disentangle the role of condensate dynamics from that of input geometry alone.

After this physical or computational transformation stage, the final mapping to the handwritten target domain is learned by a multilayer perceptron (MLP) generator. To train this mapping, we employ a conditional Wasserstein GAN with gradient penalty (WGAN-GP) framework~\cite{gulrajani2017improved,arjovsky2017wasserstein}, which provides robust adversarial training based on the Wasserstein distance.

Given an input image $\mathbf{x}_{\text{input}}$ and its corresponding digit label $y \in \{0, 1\}$, the generator $G$ learns to produce a handwritten-style output $\mathbf{x}_{\text{output}} = G(\mathbf{x}_{\text{input}}, y)$ that preserves the digit identity while matching the visual characteristics of handwritten MNIST digits~\cite{lecun1998gradient}. The input is taken either as a few-shot measurement of a photoluminescence/reflection pattern arising from nonlinear polariton dynamics (condensate mode, \raisebox{.5pt}{\textcircled{\raisebox{-.9pt} {1}}}) or as a standard Helvetica digit with perturbations added numerically (baseline mode, \raisebox{.5pt}{\textcircled{\raisebox{-.9pt} {2}}}).
Figure~\ref{fig:fig1} illustrates this dual-mode approach: condensate processing (green arrows) uses the nonlinear dynamics and stochastic fluctuations of polariton condensates to transform input digits, while the computational baseline (purple arrows) uses random perturbation injection. Both modes feed into the same WGAN-GP architecture for adversarial training. The architecture consists of three main components: (1) a three-layer multi-layer perceptron (MLP) generator $G$ that transforms input images conditioned on class labels, (2) a three-layer MLP critic $D$ that evaluates the authenticity of generated samples, and (3) a convolutional neural network (CNN) classifier $C$ for digit preservation evaluation. The detailed architecture specifications of the generator, critic, and classifier networks, are provided in Methods.

The training follows a minimax optimization where the critic $D$ aims to maximize the Wasserstein distance between generated and real distributions
\begin{equation}
\mathcal{L}_D = \mathbb{E}_{\mathbf{x}\sim p_{\text{real}}}[D(\mathbf{x}, y)] - \mathbb{E}_{\mathbf{x}\sim p_G}[D(G(\mathbf{x}_{\text{input}}, y), y)]
- \lambda_{\mathrm{GP}} \cdot \text{GP},
\end{equation}
where $p_{\text{real}}$ denotes the distribution of real MNIST samples, $p_G$ denotes the generator distribution, and the gradient penalty term $\text{GP} = \mathbb{E}_{\hat{\mathbf{x}}}[(\|\nabla_{\hat{\mathbf{x}}} D(\hat{\mathbf{x}}, y)\|_2 - 1)^2]$ enforces the Lipschitz constraint with $\hat{\mathbf{x}} = \epsilon \mathbf{x}_{\text{real}} + (1-\epsilon) \mathbf{x}_{\text{fake}}$ and $\epsilon \sim \text{Uniform}(0,1)$~\cite{gulrajani2017improved}. Here, we set $\lambda_{\mathrm{GP}} = 10$. The generator minimizes
\begin{equation}
\mathcal{L}_G = -\mathbb{E}_{\mathbf{x}\sim p_G}[D(G(\mathbf{x}_{\text{input}}, y), y)],
\end{equation}
driving the generated samples toward the real data distribution through this adversarial process.

The fundamental distinction between our two experimental modes lies in the origin of input variability: in the condensate mode, inputs are transformed through the nonlinear many-body dynamics of the polariton condensate, which produces spatially-correlated intensity variations governed by coherent polariton-polariton interactions, whereas the baseline mode employs a random number generator to add uncorrelated perturbations to standard Helvetica digits. This comparison allows us to assess whether the structured variability arising from nonlinear condensate dynamics offers computational advantages over purely digital random perturbation generation. A pre-trained CNN classifier $C$ is used solely for evaluation to assess digit identity preservation but does not contribute to the adversarial training procedure itself.

\subsection{Experimental realization of Polaritonic GAN}%added by Marcin

The experimental realization of the polariton image-processing module is illustrated in Figure~\ref{fig:fig2}(a) (see Supplementary Note 5 for technical details of the setup and sample). A femtosecond laser beam is directed onto a reflective SLM operated in phase-only mode, generating a holographic image of the Helvetica numeral `$0$' or `$1$' in the back focal plane of a microscope objective. The objective projects the shaped excitation profile onto the optical microcavity sample, where an nBEC of polaritons is formed. The resulting PL follows the spatial profile of the excitation, modified by the nonlinear spatiotemporal response of the condensate. Both the reflected pump beam and the PL signal are directed by a beam splitter towards a CCD camera, while a long-pass filter suppresses the reflected pump. The sample used in this work is a Tamm plasmon optical microcavity filled with R3B-SMILES active medium, operating in the strong-coupling regime and supporting polariton condensation, as reported in Ref.~\cite{deshmukh2024plug}.
\begin{figure*}[h]
\begin{center}
\includegraphics[scale=0.5]{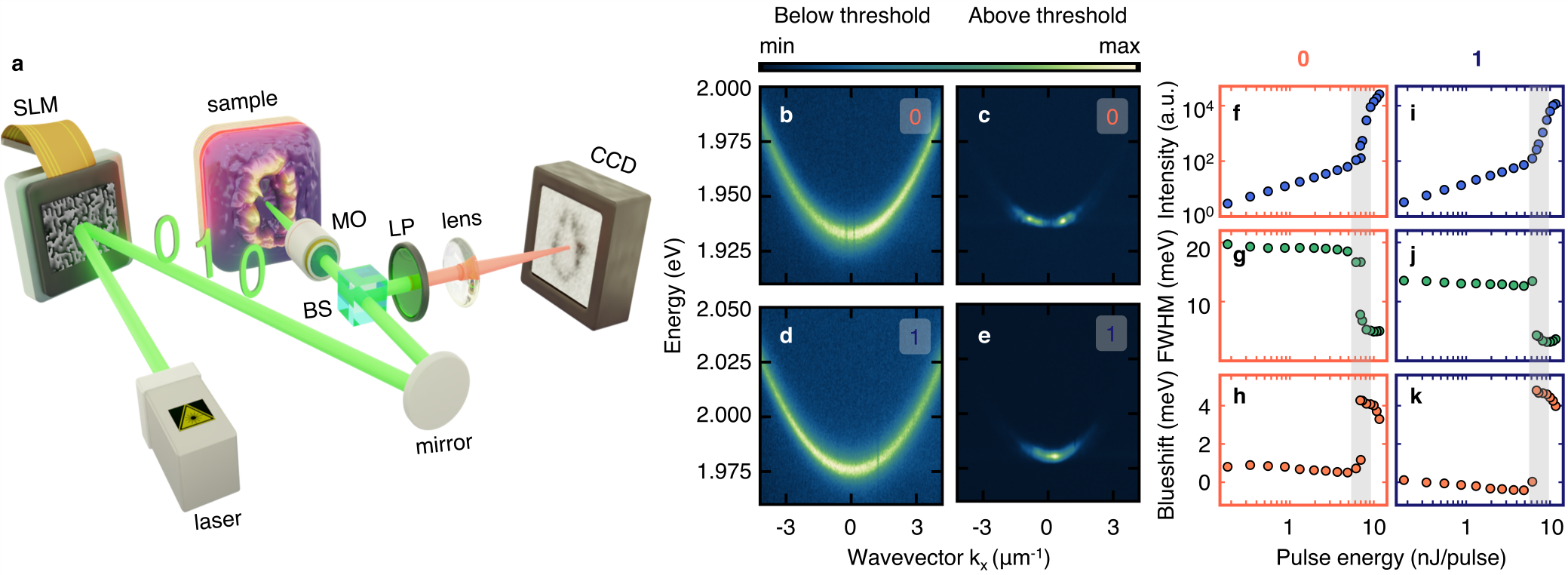}
\end{center}
\caption{\textbf{Experimental realization of polaritonic generative modeling.} (a) Scheme of the experimental setup for nonlinear image transformation. SLM, spatial light modulator; MO, microscope objective; BS, beam splitter; LP, longpass filter; CCD, charge-coupled device camera. (b--e) Normalized momentum-resolved photoluminescence spectra collected (b,d) below and (c,e) above the condensation threshold for excitation with a `$0$' (b,c) and a `$1$' (d,e) laser intensity profile. (f,i) Integrated photoluminescence intensity, (g,j) full width at half maximum, and (h,k) emission blueshift as a function of excitation pulse energy for the `$0$' (f--h) and `$1$' (i--k) excitation profiles. Gray shaded regions indicate the condensation threshold, our regime of operation.}
\label{fig:fig2}
\end{figure*}

The PL spectrum of the nBEC strongly depends on the excitation pulse energy, as illustrated by the representative momentum-resolved spectra in Fig.~\ref{fig:fig2}(b)-(e), collected below and above the condensation threshold for `$0$' and `$1$' excitation profiles. Below threshold, the emission in both cases is distributed along the lower polariton branch [Fig.~\ref{fig:fig2}(b,d)]. Above threshold, it becomes localized near the minimum of the lower polariton dispersion [Fig.~\ref{fig:fig2}(c,e)], indicating macroscopic occupation of a single polariton state. For the `$0$' excitation profile, which effectively acts as a polariton trap, the condensate reaches its maximum intensity at finite in-plane momentum, a characteristic feature of trapped condensates~\cite{estrecho2019direct}. Despite these qualitative differences, the overall nonlinear response remains similar for both excitation profiles, as evidenced by the threshold behaviour of the emission intensity [Fig.~\ref{fig:fig2}(f,i)], linewidth narrowing [Fig.~\ref{fig:fig2}(g,j)], and the condensation-induced blueshift of approximately 4~meV [Fig.~\ref{fig:fig2}(h,k)], typical for organic polariton condensates~\cite{daskalakis2014nonlinear}.

The PL signal below threshold is nearly undetectable in a near-single-shot regime. To achieve a sufficient signal-to-noise ratio, spectra below threshold were accumulated over $120{,}000$ excitation pulses, whereas above-threshold spectra were collected using 200 pulses. To prepare the training dataset for the polariton-enriched GAN architecture (condensate mode), all data were acquired slightly above the condensation threshold using five pulses per frame, providing a practical compromise between signal-to-noise ratio and the shot-to-shot variability required by the platform (see Supplementary Note 5 and Fig.~S1 for details).

\subsection{Training dynamics and model selection}

\begin{figure*}[h]
\begin{center}
\includegraphics[scale=1]{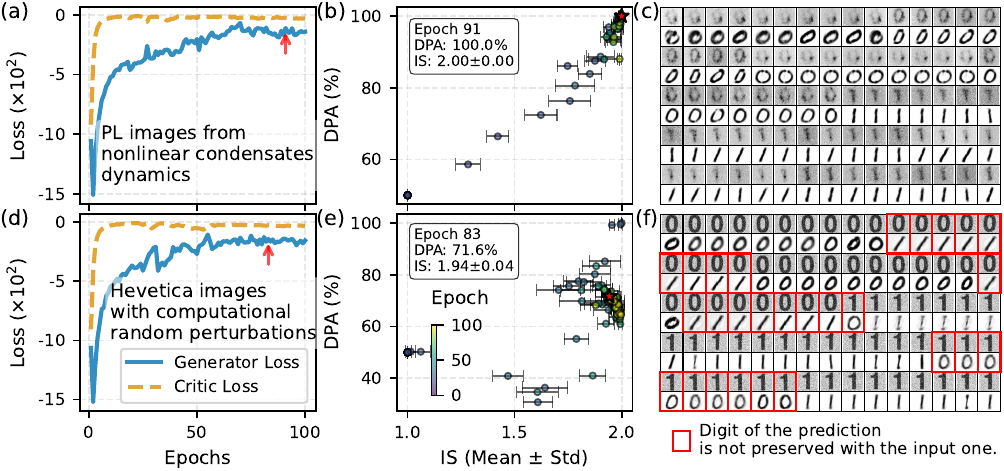}
\end{center}
\caption{\textbf{Training dynamics and performance of condensate and baseline conditional GAN modes.} (a,\,d) Generator and critic loss evolution over $100$ training epochs for (a) condensate mode using PL images generated through nonlinear condensate dynamics and (d) baseline mode using standard Helvetica digits with computational random perturbations. Red arrows indicate best epochs selected for the model of the generator (epoch $91$ and $83$, respectively). (b,\,e) Digit preservation accuracy (DPA) versus inception score (IS) across all training epochs for (b) condensate mode and (e) baseline mode. Each marker represents model at one epoch shown with different color, with error bars showing IS standard deviation across splits. Red stars mark the selected epochs based on the highest IS $\times$ DPA product among non-collapsed models. (c,\,f)~Representative grids of generated MNIST-style digit images from (c) the condensate model and (f) the baseline model. In each grid, rows alternate between input images (odd rows) and their corresponding generated predictions (even rows), forming paired input-output examples. Red rectangles in (f) highlight cases where digit identity is not preserved during translation.}
\label{fig:fig3}
\end{figure*}

Fig.~\ref{fig:fig3} presents the training dynamics and performance comparison between the two modes. Figs.~\ref{fig:fig3}~(a) and~(d) show the evolution of generator and critic losses over $100$ training epochs for the condensate and baseline modes, respectively. In adversarial training, these losses do not converge to zero but rather reach a dynamic equilibrium~\cite{goodfellow2016nips}: when the generator produces realistic samples that successfully fool the critic, the generator loss decreases while the critic loss increases as it struggles to distinguish real data from generated data. However, loss values alone do not reveal the full picture of model performance. A fundamental challenge in generative modeling is mode collapse~\cite{goodfellow2016nips,salimans2016improved}, where the generator learns to produce only a limited subset of plausible outputs rather than capturing the full diversity of the target distribution. Such collapsed models can achieve deceptively good loss values yet fail the fundamental requirement of preserving digit identity.
%%%
%%%

To detect and quantify these failure modes, we evaluate each epoch's model using two complementary metrics: inception score (IS)~\cite{salimans2016improved} and digit preservation accuracy (DPA), where precise definitions included in Supplementary Note 4.  The IS, adapted from the standard formulation for our binary-class task, measures both the quality and diversity of generated samples through the divergence between conditional and marginal class distributions predicted by the classifier $C$. For binary classification, IS ranges from $1.0$ to $2.0$. The DPA directly measures the percentage of generated images correctly classified as their intended digit by the pre-trained classifier. It is crucial to recognize that neither metric alone, nor their product (IS $\times$ DPA), provides a complete assessment of model quality. A model might achieve high DPA by consistently generating digit-like patterns that are correctly classified but visually indistinct or unrealistic (low IS). Conversely, a model could produce highly realistic and diverse handwritten digits (high IS) that do not correspond to the input labels (low DPA). The product IS $\times$ DPA attempts to balance these objectives but cannot capture all aspects of generation quality. For instance, it cannot distinguish between visually realistic outputs and those that merely satisfy the classifier.

\section{RESULTS}\label{sec-res}

We evaluated the performance of conditional WGAN-GP models across four independent random initializations to assess robustness and consistency. The condensate mode uses the nonlinear dynamics and stochastic fluctuations inherent to polariton condensate formation, while the baseline mode employs computationally-generated Gaussian perturbations detailed in Supplementary~Note~2. Additionally, we test a lasing mode that uses below-threshold patterns augmented with computational perturbations. All models were trained using identical architectures and hyperparameters (generator learning rate $l_G = 10^{-4}$, critic learning rate $l_C = 5 \times 10^{-5}$, batch size $8$), enabling direct comparison of how different sources of input variability affect generative performance. Each model was trained for $100$ epochs, with the best-performing checkpoint selected based on validation metrics.

Figure~\ref{fig:fig3} presents training dynamics and performance comparison for one representative seed. Fig.~\ref{fig:fig3}(a,\,d) display the evolution of generator and critic losses over training. Both modes exhibit the characteristic adversarial equilibrium where losses fluctuate rather than converge to zero~\cite{goodfellow2016nips}. To assess model quality, we evaluate each epoch using DPA and IS. Fig.~\ref{fig:fig3}(b,\,e) show these metrics across all training epochs. The condensate model [Fig.~\ref{fig:fig3}(b)] demonstrates a clear progression toward the optimal upper-right region (high DPA, high IS), achieving perfect digit preservation (DPA = $100.0\%$) with maximum IS = $2.00$ at epoch $91$. Conversely, the baseline model [Fig.~\ref{fig:fig3}~(e)] struggles to simultaneously achieve high DPA and high IS, with the best-performing epoch (epoch $83$) reaching only DPA = $71.6\%$ and IS = $1.94$. This fundamental trade-off in the baseline model reflects mode collapse tendencies where the generator fails to capture the full diversity of target handwritten digits while preserving digit identity. The spatial clustering of misclassified outputs in Fig.~\ref{fig:fig3}(f) is characteristic of mode collapse in GAN training~\cite{goodfellow2016nips,salimans2016improved}. Rather than randomly distributed failures, the generator learns specific output patterns that successfully fool the critic despite failing to preserve digit identity. This behavior, as shown in Fig.~\ref{fig:fig3}(f) where the red rectangles highlight multiple `0' inputs incorrectly mapping to `1' patterns, indicates the generator exploited a suboptimal solution space. The computational perturbations lack spatial correlations, allowing the generator to achieve locally optimal critic scores through repetitive, semantically-incorrect outputs. In contrast, the condensate model [see Fig.~\ref{fig:fig3}(c)] exhibits no such clustering, as the spatially-correlated fluctuations arising from
nonlinear polariton interactions provide natural constraints that guide the generator toward digit-identity-preserving transformations. Representative generation grids [Fig.~\ref{fig:fig3}(c,\,f)] visualize this performance gap. The condensate model [Fig.~\ref{fig:fig3}(c)] successfully translates both digit classes while maintaining structural coherence and digit identity. The baseline model [Fig.~\ref{fig:fig3}(f)] exhibits misclassified failures, indicating the model learned incomplete mappings from input to target distributions.

We evaluate the best-performing models from all four seeds on independent test sets containing $2,304$ samples each ($1,152$ per digit class). Table~\ref{tab:results} summarizes the comprehensive performance comparison across all metrics. The condensate model achieves perfect digit preservation ($100.00\%$ overall across all four seeds) with perfect consistency across all random initializations, demonstrating exceptional robustness. The laser model achieves $99.89^{+0.02}_{-0.07}\%$ mean accuracy. In contrast, the baseline model achieves $94.02^{+5.98}_{-17.72}\%$ mean overall accuracy with wide variability. This large asymmetric error reflects the baseline's susceptibility to mode collapse.
%%%
\begin{table}[h]
\centering
\caption{Performance comparison between condensate mode with nonlinear condensate dynamics (NCD), reflected laser with computational perturbations (CP) and baseline with CP GAN modes on test sets, averaged across four independent random initializations. All metrics computed on $2,304$ test samples with equal class balance. For DPA metrics, we report mean with asymmetric errors: mean$^{+\text{upper}}_{-\text{lower}}$ where upper (lower) indicate deviations to maximum (minimum) observed values; for other metrics, mean $\pm$ standard deviation. DPA: digit preservation accuracy, higher is better; IS: inception score, higher is better; SSIM: structural similarity index (pairwise among generated outputs), lower is better. Pixel var: average variance of pixel intensities.}
\label{tab:results}
\begin{tabular}{lccc}
\hline\hline
\text{Metric} & \text{Condensate (NCD)} & \text {Laser (CP)} & \text{Baseline (CP)} \\
DPA (\%) & \textbf{$100.00^{+0.00}_{-0.00}$} & \text{$99.89^{+0.02}_{-0.07}$} & $94.02^{+5.98}_{-17.72}$ 
\\
IS & 1.9987 $\pm$ 0.0009 & 1.9992 $\pm$ 0.0003  & 1.9727 $\pm$ 0.0432 
\\
SSIM & 0.6839 $\pm$ 0.0198 & 0.7915 $\pm$ 0.0155  & 0.6970 $\pm$ 0.0264 
\\
Pixel var. & 3083 $\pm$ 298 & 1732 $\pm$ 225 & 2783 $\pm$ 178 \\
\hline\hline
\end{tabular}
\end{table}
%%%

All three models achieve near-theoretical maximum IS values for binary classification. As shown in Table~\ref{tab:results}, these values indicate that generated samples are: (1) confidently classified by the pre-trained classifier ($p(y|\mathbf{x})$ has low entropy), and (2) uniformly distributed across both classes ($p(y)$ is balanced). For cases where all models achieved comparable DPA ($\sim$100\%), IS alone cannot distinguish generation quality differences. This limitation motivates the critical importance of our third metric: pairwise structural similarity (SSIM) among generated outputs.

To assess output diversity and to understand whether models generate varied samples rather than memorizing or producing nearly identical outputs, we compute pairwise SSIM among all generated images within each test set. Crucially, lower pairwise SSIM indicates higher diversity (generated images are less similar to each other), while higher pairwise SSIM suggests the model produces repetitive outputs. The condensate mode exhibits the lowest pairwise SSIM ($0.6839 \pm 0.0198$), compared to the laser mode ($0.7915 \pm 0.0155$) and baseline mode ($0.6970 \pm 0.0264$), demonstrating that polariton-mediated nonlinear transformations generate more diverse outputs. The baseline model’s intermediate SSIM performance, despite having the lowest DPA, indicates it generates varied outputs, but many are incorrect, reflecting mode collapse to semantically-incorrect yet diverse patterns. The pixel-level variance analysis further supports these findings: the polaritonic model exhibits the highest pixel variance ($3083 \pm 298$), compared to the laser model ($1732 \pm 225$) and the baseline ($2783 \pm 178$), indicating greater variability in spatial intensity patterns across generated samples. The consistency of the polaritonic model advantages across all independent initializations demonstrates that the diversity and stability improvements are shared properties for nonlinear condensate mode rather than a statistical artifact.

\section{DISCUSSION}

\begin{figure*}[h]
\begin{center}
\includegraphics[scale=1]{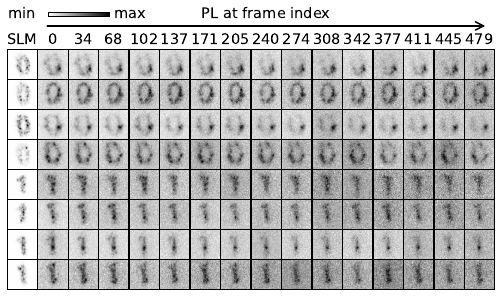}
\end{center}
\caption{\textbf{Evolution of polariton condensate photoluminescence across successive experimental realizations.} The first column shows the spatial light modulator (SLM) images used to generate different excitation configurations. Each row displays the corresponding condensate PL intensity at selected realization indices from $0$ to $479$ (out of $480$ total realizations), with evenly spaced sampling showing the shot-to-shot variation. The eight rows, with the first four corresponding to class `0' and the remaining four to class `1', demonstrate different initial pattern geometries and their evolution, revealing condensate formation dynamics and spatial redistribution across realizations.}
\label{fig:fig4}
\end{figure*}

The physical origin of the observed performance advantages lies in the nonlinear many-body dynamics inherent to polariton condensate formation and evolution. As shown in Fig.~\ref{fig:fig4}, the evolution of the reflectivity/PL scans across successive experimental realizations reveals the complex, nonlinear transformation process that distinguishes our polaritonic approach from conventional perturbation injection. Each row in Fig.~\ref{fig:fig4} demonstrates how identical SLM excitation configurations (first column) give rise to diverse image realizations across different shots, encoding structured variability through the condensate’s nonlinear dynamics and stochastic condensation process. This shot-to-shot variation, visible across all $480$ captured
images per SLM configuration, arises from the intrinsic stochasticity of polariton condensate formation combined with nonlinear polariton-polariton interactions, providing the physically-generated structured variability that stabilizes
GAN training.
%%%
%%%

Unlike the uncorrelated Gaussian perturbations employed in the baseline mode, the polariton-generated transformations shown in Fig.~\ref{fig:fig4} exhibit spatial correlations governed by characteristic length scales determined by the polariton effective mass and nonlinear interaction strength~\cite{deng2010exciton,estrecho2019direct}. The coherent many-body dynamics of the condensate introduce correlations both within individual realizations and across the sequence of shots, a direct consequence of the macroscopic quantum coherence and nonlinear self-interactions that characterize the Bose-Einstein condensate state. These structured variations, fundamentally distinct from numerical random noise generation, provide a dual benefit for GAN tasks: (1) they reduce the effective dimensionality of the optimization landscape through physically-correlated input variations, thereby stabilizing training against mode collapse, and (2) they naturally promote intra-class diversity by exploring valid stylistic variations while respecting the topological and geometric constraints imposed by digit identity through the spatial coherence of the condensate. A possibly related role of noise engineering was reported for quantum optical GANs~\cite{bacarreza2025quantum}.

A central question remains: can similar performance be achieved by combining the SLM spatial patterns shown in Fig.~\ref{fig:fig4} with computational random perturbations, as prepared for the baseline model? To address this question, we prepared datasets consisting of the same number of the condensate PL images but with computational perturbations added to laser-generated PL images which is termed the laser mode. The laser mode results shown in Table~\ref{tab:results} provide critical insights into disentangling the contributions of physical spatial structure versus nonlinear condensate dynamics. While the laser-generated SLM images possess rich spatial characteristics similar to polariton condensate-generated patterns, they lack the stochasticity and nonlinear transformations arising from the coherent many-body dynamics of the polaritonic system. The laser mode's near-perfect DPA and high IS initially suggest excellent performance. However, deeper analysis reveals a subtle failure mode: the laser mode exhibits significantly higher pairwise SSIM ($0.7915 \pm 0.0155$) compared to the condensate mode ($0.6839 \pm 0.0198$), indicating substantially reduced output diversity. Also, the $78\%$ lower pixel variance in laser outputs ($1732 \pm 225$) compared to condensate outputs ($3083 \pm 298$) quantifies this reduced exploration of the output manifold. This collapse to a narrower output distribution occurs because computational perturbation augmentation, even when applied to realistic SLM spatial patterns, lacks the spatially-correlated, temporally-evolving variations that emerge directly from polariton condensate dynamics. The generator learns to map laser images to a narrow subset of valid MNIST-style outputs that successfully preserve digit identity and fool the critic (achieving high DPA and IS), but fail to capture the full stylistic diversity present in handwritten digits. This failure is invisible to standard GAN metrics (DPA and IS) which focus on semantic correctness and classifier confidence, highlighting the critical importance of diversity metrics (SSIM and pixel variance) in detecting mode collapse. The laser mode thus demonstrates that achieving realistic spatial structure alone without the structured variability arising from nonlinear polariton dynamics leads to incomplete generative performance.

The interpretation of variability in our hybrid architecture differs fundamentally from traditional GAN formulations. Rather than sampling from a learned latent distribution~\cite{goodfellow2014generative}, the input digit patterns serve as structured physical latent representations encoding digit identity, while the polariton condensate acts as a physics-based stochastic nonlinear layer that transforms these inputs during the condensate formation. Crucially, the polariton system does not learn a mapping, but instead provides a physical transformation whose output variations are governed by the condensate's nonlinear response, polariton-polariton interaction strength and coherence properties of the macroscopic quantum state. The conditional GAN then learns to translate these physically-transformed patterns (see the diverse PL outputs in Fig.~\ref{fig:fig4}) into the target MNIST domain. This positions polariton condensate dynamics as physics-based data augmentation operating in continuous physical space through coherent many-body interactions, fundamentally distinguishing our approach from purely computational random perturbation methods.

\section{CONCLUSION}

We have demonstrated conditional generative modelling with room-temperature exciton-polariton condensates acting as a physical stochastic nonlinear layer within a hybrid GAN architecture. In digit-to-image translation, the condensate mode outperforms digital and laser-based baseline models, yielding stronger digit preservation together with greater output diversity. The key advantage is that driven nonlinear many-body dynamics of the physical layer generate structured variability, stabilizing adversarial training and supporting richer exploration of the target distribution.

Our results position polariton condensation as a useful computational resource for generative modelling. More broadly, they point to a class of nonlinear photonic hardware in which stochastic physical dynamics contribute directly to generation, rather than serving only as preprocessing or inference layers. This opens a route towards physics-enhanced generative machine learning based on strongly coupled light-matter systems operating at room temperature.

\section{Data availability}

Data will be
made available in a repository after acceptance of the final version.

\section{Code availability}

Code will be
made available in a repository after acceptance of the final version.
\\
\begin{acknowledgments}
\textit{Acknowledgments.} O.K. and Y.W. acknowledge the support from UK EPSRC grants EP/X017222/1 and EP/Z53318X/1. A.D and V.M.M were supported by US Army Research Office grant W911NF-22-1-0091 and M.M was suported by National Science Foundation grant OMA 2328993.
\end{acknowledgments}

%\bibliography{main}

\begin{thebibliography}{67}%
\makeatletter
\providecommand \@ifxundefined [1]{%
 \@ifx{#1\undefined}
}%
\providecommand \@ifnum [1]{%
 \ifnum #1\expandafter \@firstoftwo
 \else \expandafter \@secondoftwo
 \fi
}%
\providecommand \@ifx [1]{%
 \ifx #1\expandafter \@firstoftwo
 \else \expandafter \@secondoftwo
 \fi
}%
\providecommand \natexlab [1]{#1}%
\providecommand \enquote  [1]{``#1''}%
\providecommand \bibnamefont  [1]{#1}%
\providecommand \bibfnamefont [1]{#1}%
\providecommand \citenamefont [1]{#1}%
\providecommand \href@noop [0]{\@secondoftwo}%
\providecommand \href [0]{\begingroup \@sanitize@url \@href}%
\providecommand \@href[1]{\@@startlink{#1}\@@href}%
\providecommand \@@href[1]{\endgroup#1\@@endlink}%
\providecommand \@sanitize@url [0]{\catcode `\\12\catcode `\$12\catcode `\&12\catcode `\#12\catcode `\^12\catcode `\_12\catcode `\%12\relax}%
\providecommand \@@startlink[1]{}%
\providecommand \@@endlink[0]{}%
\providecommand \url  [0]{\begingroup\@sanitize@url \@url }%
\providecommand \@url [1]{\endgroup\@href {#1}{\urlprefix }}%
\providecommand \urlprefix  [0]{URL }%
\providecommand \Eprint [0]{\href }%
\providecommand \doibase [0]{https://doi.org/}%
\providecommand \selectlanguage [0]{\@gobble}%
\providecommand \bibinfo  [0]{\@secondoftwo}%
\providecommand \bibfield  [0]{\@secondoftwo}%
\providecommand \translation [1]{[#1]}%
\providecommand \BibitemOpen [0]{}%
\providecommand \bibitemStop [0]{}%
\providecommand \bibitemNoStop [0]{.\EOS\space}%
\providecommand \EOS [0]{\spacefactor3000\relax}%
\providecommand \BibitemShut  [1]{\csname bibitem#1\endcsname}%
\let\auto@bib@innerbib\@empty
%</preamble>
\bibitem [{\citenamefont {Goodfellow}\ \emph {et~al.}(2014)\citenamefont {Goodfellow}, \citenamefont {Pouget-Abadie}, \citenamefont {Mirza}, \citenamefont {Xu}, \citenamefont {Warde-Farley}, \citenamefont {Ozair}, \citenamefont {Courville},\ and\ \citenamefont {Bengio}}]{goodfellow2014generative}%
  \BibitemOpen
  \bibfield  {author} {\bibinfo {author} {\bibfnamefont {I.~J.}\ \bibnamefont {Goodfellow}}, \bibinfo {author} {\bibfnamefont {J.}~\bibnamefont {Pouget-Abadie}}, \bibinfo {author} {\bibfnamefont {M.}~\bibnamefont {Mirza}}, \bibinfo {author} {\bibfnamefont {B.}~\bibnamefont {Xu}}, \bibinfo {author} {\bibfnamefont {D.}~\bibnamefont {Warde-Farley}}, \bibinfo {author} {\bibfnamefont {S.}~\bibnamefont {Ozair}}, \bibinfo {author} {\bibfnamefont {A.}~\bibnamefont {Courville}},\ and\ \bibinfo {author} {\bibfnamefont {Y.}~\bibnamefont {Bengio}},\ }\bibfield  {title} {\bibinfo {title} {Generative adversarial nets},\ }\href@noop {} {\bibfield  {journal} {\bibinfo  {journal} {Advances in neural information processing systems}\ }\textbf {\bibinfo {volume} {27}} (\bibinfo {year} {2014})}\BibitemShut {NoStop}%
\bibitem [{\citenamefont {Ho}\ \emph {et~al.}(2020)\citenamefont {Ho}, \citenamefont {Jain},\ and\ \citenamefont {Abbeel}}]{ho2020denoising}%
  \BibitemOpen
  \bibfield  {author} {\bibinfo {author} {\bibfnamefont {J.}~\bibnamefont {Ho}}, \bibinfo {author} {\bibfnamefont {A.}~\bibnamefont {Jain}},\ and\ \bibinfo {author} {\bibfnamefont {P.}~\bibnamefont {Abbeel}},\ }\bibfield  {title} {\bibinfo {title} {Denoising diffusion probabilistic models},\ }\href@noop {} {\bibfield  {journal} {\bibinfo  {journal} {Advances in neural information processing systems}\ }\textbf {\bibinfo {volume} {33}},\ \bibinfo {pages} {6840} (\bibinfo {year} {2020})}\BibitemShut {NoStop}%
\bibitem [{\citenamefont {Vaswani}\ \emph {et~al.}(2017)\citenamefont {Vaswani}, \citenamefont {Shazeer}, \citenamefont {Parmar}, \citenamefont {Uszkoreit}, \citenamefont {Jones}, \citenamefont {Gomez}, \citenamefont {Kaiser},\ and\ \citenamefont {Polosukhin}}]{Vaswani2017attention}%
  \BibitemOpen
  \bibfield  {author} {\bibinfo {author} {\bibfnamefont {A.}~\bibnamefont {Vaswani}}, \bibinfo {author} {\bibfnamefont {N.}~\bibnamefont {Shazeer}}, \bibinfo {author} {\bibfnamefont {N.}~\bibnamefont {Parmar}}, \bibinfo {author} {\bibfnamefont {J.}~\bibnamefont {Uszkoreit}}, \bibinfo {author} {\bibfnamefont {L.}~\bibnamefont {Jones}}, \bibinfo {author} {\bibfnamefont {A.~N.}\ \bibnamefont {Gomez}}, \bibinfo {author} {\bibfnamefont {{\L}.}~\bibnamefont {Kaiser}},\ and\ \bibinfo {author} {\bibfnamefont {I.}~\bibnamefont {Polosukhin}},\ }\bibfield  {title} {\bibinfo {title} {Attention is all you need},\ }\href@noop {} {\bibfield  {journal} {\bibinfo  {journal} {Advances in neural information processing systems}\ }\textbf {\bibinfo {volume} {30}} (\bibinfo {year} {2017})}\BibitemShut {NoStop}%
\bibitem [{\citenamefont {Brown}\ \emph {et~al.}(2020)\citenamefont {Brown}, \citenamefont {Mann}, \citenamefont {Ryder}, \citenamefont {Subbiah}, \citenamefont {Kaplan}, \citenamefont {Dhariwal}, \citenamefont {Neelakantan}, \citenamefont {Shyam}, \citenamefont {Sastry}, \citenamefont {Askell} \emph {et~al.}}]{brown2020language}%
  \BibitemOpen
  \bibfield  {author} {\bibinfo {author} {\bibfnamefont {T.}~\bibnamefont {Brown}}, \bibinfo {author} {\bibfnamefont {B.}~\bibnamefont {Mann}}, \bibinfo {author} {\bibfnamefont {N.}~\bibnamefont {Ryder}}, \bibinfo {author} {\bibfnamefont {M.}~\bibnamefont {Subbiah}}, \bibinfo {author} {\bibfnamefont {J.~D.}\ \bibnamefont {Kaplan}}, \bibinfo {author} {\bibfnamefont {P.}~\bibnamefont {Dhariwal}}, \bibinfo {author} {\bibfnamefont {A.}~\bibnamefont {Neelakantan}}, \bibinfo {author} {\bibfnamefont {P.}~\bibnamefont {Shyam}}, \bibinfo {author} {\bibfnamefont {G.}~\bibnamefont {Sastry}}, \bibinfo {author} {\bibfnamefont {A.}~\bibnamefont {Askell}}, \emph {et~al.},\ }\bibfield  {title} {\bibinfo {title} {Language models are few-shot learners},\ }\href@noop {} {\bibfield  {journal} {\bibinfo  {journal} {Advances in neural information processing systems}\ }\textbf {\bibinfo {volume} {33}},\ \bibinfo {pages} {1877} (\bibinfo {year} {2020})}\BibitemShut {NoStop}%
\bibitem [{\citenamefont {Sanchez-Lengeling}\ and\ \citenamefont {Aspuru-Guzik}(2018)}]{sanchez2018inverse}%
  \BibitemOpen
  \bibfield  {author} {\bibinfo {author} {\bibfnamefont {B.}~\bibnamefont {Sanchez-Lengeling}}\ and\ \bibinfo {author} {\bibfnamefont {A.}~\bibnamefont {Aspuru-Guzik}},\ }\bibfield  {title} {\bibinfo {title} {Inverse molecular design using machine learning: Generative models for matter engineering},\ }\href@noop {} {\bibfield  {journal} {\bibinfo  {journal} {Science}\ }\textbf {\bibinfo {volume} {361}},\ \bibinfo {pages} {360} (\bibinfo {year} {2018})}\BibitemShut {NoStop}%
\bibitem [{\citenamefont {Yao}\ \emph {et~al.}(2021)\citenamefont {Yao}, \citenamefont {S{\'a}nchez-Lengeling}, \citenamefont {Bobbitt}, \citenamefont {Bucior}, \citenamefont {Kumar}, \citenamefont {Collins}, \citenamefont {Burns}, \citenamefont {Woo}, \citenamefont {Farha}, \citenamefont {Snurr} \emph {et~al.}}]{yao2021inverse}%
  \BibitemOpen
  \bibfield  {author} {\bibinfo {author} {\bibfnamefont {Z.}~\bibnamefont {Yao}}, \bibinfo {author} {\bibfnamefont {B.}~\bibnamefont {S{\'a}nchez-Lengeling}}, \bibinfo {author} {\bibfnamefont {N.~S.}\ \bibnamefont {Bobbitt}}, \bibinfo {author} {\bibfnamefont {B.~J.}\ \bibnamefont {Bucior}}, \bibinfo {author} {\bibfnamefont {S.~G.~H.}\ \bibnamefont {Kumar}}, \bibinfo {author} {\bibfnamefont {S.~P.}\ \bibnamefont {Collins}}, \bibinfo {author} {\bibfnamefont {T.}~\bibnamefont {Burns}}, \bibinfo {author} {\bibfnamefont {T.~K.}\ \bibnamefont {Woo}}, \bibinfo {author} {\bibfnamefont {O.~K.}\ \bibnamefont {Farha}}, \bibinfo {author} {\bibfnamefont {R.~Q.}\ \bibnamefont {Snurr}}, \emph {et~al.},\ }\bibfield  {title} {\bibinfo {title} {Inverse design of nanoporous crystalline reticular materials with deep generative models},\ }\href@noop {} {\bibfield  {journal} {\bibinfo  {journal} {Nature Machine Intelligence}\ }\textbf {\bibinfo {volume} {3}},\ \bibinfo {pages} {76} (\bibinfo {year} {2021})}\BibitemShut {NoStop}%
\bibitem [{\citenamefont {Wang}\ \emph {et~al.}(2023{\natexlab{a}})\citenamefont {Wang}, \citenamefont {Fu}, \citenamefont {Du}, \citenamefont {Gao}, \citenamefont {Huang}, \citenamefont {Liu}, \citenamefont {Chandak}, \citenamefont {Liu}, \citenamefont {Van~Katwyk}, \citenamefont {Deac} \emph {et~al.}}]{wang2023scientific}%
  \BibitemOpen
  \bibfield  {author} {\bibinfo {author} {\bibfnamefont {H.}~\bibnamefont {Wang}}, \bibinfo {author} {\bibfnamefont {T.}~\bibnamefont {Fu}}, \bibinfo {author} {\bibfnamefont {Y.}~\bibnamefont {Du}}, \bibinfo {author} {\bibfnamefont {W.}~\bibnamefont {Gao}}, \bibinfo {author} {\bibfnamefont {K.}~\bibnamefont {Huang}}, \bibinfo {author} {\bibfnamefont {Z.}~\bibnamefont {Liu}}, \bibinfo {author} {\bibfnamefont {P.}~\bibnamefont {Chandak}}, \bibinfo {author} {\bibfnamefont {S.}~\bibnamefont {Liu}}, \bibinfo {author} {\bibfnamefont {P.}~\bibnamefont {Van~Katwyk}}, \bibinfo {author} {\bibfnamefont {A.}~\bibnamefont {Deac}}, \emph {et~al.},\ }\bibfield  {title} {\bibinfo {title} {Scientific discovery in the age of artificial intelligence},\ }\href@noop {} {\bibfield  {journal} {\bibinfo  {journal} {Nature}\ }\textbf {\bibinfo {volume} {620}},\ \bibinfo {pages} {47} (\bibinfo {year} {2023}{\natexlab{a}})}\BibitemShut {NoStop}%
\bibitem [{\citenamefont {IEA}(2025)}]{IEA2025}%
  \BibitemOpen
  \bibfield  {author} {\bibinfo {author} {\bibnamefont {IEA}},\ }\href {https://www.iea.org/reports/energy-and-ai} {\emph {\bibinfo {title} {Energy and AI}}},\ \bibinfo {type} {Tech. Rep.}\ (\bibinfo  {institution} {International Energy Agency},\ \bibinfo {address} {Paris},\ \bibinfo {year} {2025})\BibitemShut {NoStop}%
\bibitem [{\citenamefont {Tanaka}\ \emph {et~al.}(2019)\citenamefont {Tanaka}, \citenamefont {Yamane}, \citenamefont {H{\'e}roux}, \citenamefont {Nakane}, \citenamefont {Kanazawa}, \citenamefont {Takeda}, \citenamefont {Numata}, \citenamefont {Nakano},\ and\ \citenamefont {Hirose}}]{tanaka2019recent}%
  \BibitemOpen
  \bibfield  {author} {\bibinfo {author} {\bibfnamefont {G.}~\bibnamefont {Tanaka}}, \bibinfo {author} {\bibfnamefont {T.}~\bibnamefont {Yamane}}, \bibinfo {author} {\bibfnamefont {J.~B.}\ \bibnamefont {H{\'e}roux}}, \bibinfo {author} {\bibfnamefont {R.}~\bibnamefont {Nakane}}, \bibinfo {author} {\bibfnamefont {N.}~\bibnamefont {Kanazawa}}, \bibinfo {author} {\bibfnamefont {S.}~\bibnamefont {Takeda}}, \bibinfo {author} {\bibfnamefont {H.}~\bibnamefont {Numata}}, \bibinfo {author} {\bibfnamefont {D.}~\bibnamefont {Nakano}},\ and\ \bibinfo {author} {\bibfnamefont {A.}~\bibnamefont {Hirose}},\ }\bibfield  {title} {\bibinfo {title} {Recent advances in physical reservoir computing: A review},\ }\href@noop {} {\bibfield  {journal} {\bibinfo  {journal} {Neural Networks}\ }\textbf {\bibinfo {volume} {115}},\ \bibinfo {pages} {100} (\bibinfo {year} {2019})}\BibitemShut {NoStop}%
\bibitem [{\citenamefont {Shastri}\ \emph {et~al.}(2021)\citenamefont {Shastri}, \citenamefont {Tait}, \citenamefont {Ferreira~de Lima}, \citenamefont {Pernice}, \citenamefont {Bhaskaran}, \citenamefont {Wright},\ and\ \citenamefont {Prucnal}}]{shastri2021photonics}%
  \BibitemOpen
  \bibfield  {author} {\bibinfo {author} {\bibfnamefont {B.~J.}\ \bibnamefont {Shastri}}, \bibinfo {author} {\bibfnamefont {A.~N.}\ \bibnamefont {Tait}}, \bibinfo {author} {\bibfnamefont {T.}~\bibnamefont {Ferreira~de Lima}}, \bibinfo {author} {\bibfnamefont {W.~H.}\ \bibnamefont {Pernice}}, \bibinfo {author} {\bibfnamefont {H.}~\bibnamefont {Bhaskaran}}, \bibinfo {author} {\bibfnamefont {C.~D.}\ \bibnamefont {Wright}},\ and\ \bibinfo {author} {\bibfnamefont {P.~R.}\ \bibnamefont {Prucnal}},\ }\bibfield  {title} {\bibinfo {title} {Photonics for artificial intelligence and neuromorphic computing},\ }\href@noop {} {\bibfield  {journal} {\bibinfo  {journal} {Nature Photonics}\ }\textbf {\bibinfo {volume} {15}},\ \bibinfo {pages} {102} (\bibinfo {year} {2021})}\BibitemShut {NoStop}%
\bibitem [{\citenamefont {McMahon}(2023)}]{mcmahon2023physics}%
  \BibitemOpen
  \bibfield  {author} {\bibinfo {author} {\bibfnamefont {P.~L.}\ \bibnamefont {McMahon}},\ }\bibfield  {title} {\bibinfo {title} {The physics of optical computing},\ }\href@noop {} {\bibfield  {journal} {\bibinfo  {journal} {Nature Reviews Physics}\ }\textbf {\bibinfo {volume} {5}},\ \bibinfo {pages} {717} (\bibinfo {year} {2023})}\BibitemShut {NoStop}%
\bibitem [{\citenamefont {Aifer}\ \emph {et~al.}(2025)\citenamefont {Aifer}, \citenamefont {Belateche}, \citenamefont {Bramhavar}, \citenamefont {Camsari}, \citenamefont {Coles}, \citenamefont {Crooks}, \citenamefont {Durian}, \citenamefont {Liu}, \citenamefont {Marchenkova}, \citenamefont {Martinez} \emph {et~al.}}]{aifer2025solving}%
  \BibitemOpen
  \bibfield  {author} {\bibinfo {author} {\bibfnamefont {M.}~\bibnamefont {Aifer}}, \bibinfo {author} {\bibfnamefont {Z.}~\bibnamefont {Belateche}}, \bibinfo {author} {\bibfnamefont {S.}~\bibnamefont {Bramhavar}}, \bibinfo {author} {\bibfnamefont {K.~Y.}\ \bibnamefont {Camsari}}, \bibinfo {author} {\bibfnamefont {P.~J.}\ \bibnamefont {Coles}}, \bibinfo {author} {\bibfnamefont {G.}~\bibnamefont {Crooks}}, \bibinfo {author} {\bibfnamefont {D.~J.}\ \bibnamefont {Durian}}, \bibinfo {author} {\bibfnamefont {A.~J.}\ \bibnamefont {Liu}}, \bibinfo {author} {\bibfnamefont {A.}~\bibnamefont {Marchenkova}}, \bibinfo {author} {\bibfnamefont {A.~J.}\ \bibnamefont {Martinez}}, \emph {et~al.},\ }\bibfield  {title} {\bibinfo {title} {Solving the compute crisis with physics-based asics},\ }\href {https://arxiv.org/abs/2507.10463} {\bibfield  {journal} {\bibinfo  {journal} {arXiv preprint arXiv:2507.10463}\ } (\bibinfo {year} {2025})}\BibitemShut {NoStop}%
\bibitem [{\citenamefont {Wetzstein}\ \emph {et~al.}(2020)\citenamefont {Wetzstein}, \citenamefont {Ozcan}, \citenamefont {Gigan}, \citenamefont {Fan}, \citenamefont {Englund}, \citenamefont {Solja{\v{c}}i{\'c}}, \citenamefont {Denz}, \citenamefont {Miller},\ and\ \citenamefont {Psaltis}}]{wetzstein2020inference}%
  \BibitemOpen
  \bibfield  {author} {\bibinfo {author} {\bibfnamefont {G.}~\bibnamefont {Wetzstein}}, \bibinfo {author} {\bibfnamefont {A.}~\bibnamefont {Ozcan}}, \bibinfo {author} {\bibfnamefont {S.}~\bibnamefont {Gigan}}, \bibinfo {author} {\bibfnamefont {S.}~\bibnamefont {Fan}}, \bibinfo {author} {\bibfnamefont {D.}~\bibnamefont {Englund}}, \bibinfo {author} {\bibfnamefont {M.}~\bibnamefont {Solja{\v{c}}i{\'c}}}, \bibinfo {author} {\bibfnamefont {C.}~\bibnamefont {Denz}}, \bibinfo {author} {\bibfnamefont {D.~A.}\ \bibnamefont {Miller}},\ and\ \bibinfo {author} {\bibfnamefont {D.}~\bibnamefont {Psaltis}},\ }\bibfield  {title} {\bibinfo {title} {Inference in artificial intelligence with deep optics and photonics},\ }\href@noop {} {\bibfield  {journal} {\bibinfo  {journal} {Nature}\ }\textbf {\bibinfo {volume} {588}},\ \bibinfo {pages} {39} (\bibinfo {year} {2020})}\BibitemShut {NoStop}%
\bibitem [{\citenamefont {Shen}\ \emph {et~al.}(2017)\citenamefont {Shen}, \citenamefont {Harris}, \citenamefont {Skirlo}, \citenamefont {Prabhu}, \citenamefont {Baehr-Jones}, \citenamefont {Hochberg}, \citenamefont {Sun}, \citenamefont {Zhao}, \citenamefont {Larochelle}, \citenamefont {Englund} \emph {et~al.}}]{shen2017deep}%
  \BibitemOpen
  \bibfield  {author} {\bibinfo {author} {\bibfnamefont {Y.}~\bibnamefont {Shen}}, \bibinfo {author} {\bibfnamefont {N.~C.}\ \bibnamefont {Harris}}, \bibinfo {author} {\bibfnamefont {S.}~\bibnamefont {Skirlo}}, \bibinfo {author} {\bibfnamefont {M.}~\bibnamefont {Prabhu}}, \bibinfo {author} {\bibfnamefont {T.}~\bibnamefont {Baehr-Jones}}, \bibinfo {author} {\bibfnamefont {M.}~\bibnamefont {Hochberg}}, \bibinfo {author} {\bibfnamefont {X.}~\bibnamefont {Sun}}, \bibinfo {author} {\bibfnamefont {S.}~\bibnamefont {Zhao}}, \bibinfo {author} {\bibfnamefont {H.}~\bibnamefont {Larochelle}}, \bibinfo {author} {\bibfnamefont {D.}~\bibnamefont {Englund}}, \emph {et~al.},\ }\bibfield  {title} {\bibinfo {title} {Deep learning with coherent nanophotonic circuits},\ }\href@noop {} {\bibfield  {journal} {\bibinfo  {journal} {Nature photonics}\ }\textbf {\bibinfo {volume} {11}},\ \bibinfo {pages} {441} (\bibinfo {year} {2017})}\BibitemShut {NoStop}%
\bibitem [{\citenamefont {Harris}\ \emph {et~al.}(2018)\citenamefont {Harris}, \citenamefont {Carolan}, \citenamefont {Bunandar}, \citenamefont {Prabhu}, \citenamefont {Hochberg}, \citenamefont {Baehr-Jones}, \citenamefont {Fanto}, \citenamefont {Smith}, \citenamefont {Tison}, \citenamefont {Alsing} \emph {et~al.}}]{harris2018linear}%
  \BibitemOpen
  \bibfield  {author} {\bibinfo {author} {\bibfnamefont {N.~C.}\ \bibnamefont {Harris}}, \bibinfo {author} {\bibfnamefont {J.}~\bibnamefont {Carolan}}, \bibinfo {author} {\bibfnamefont {D.}~\bibnamefont {Bunandar}}, \bibinfo {author} {\bibfnamefont {M.}~\bibnamefont {Prabhu}}, \bibinfo {author} {\bibfnamefont {M.}~\bibnamefont {Hochberg}}, \bibinfo {author} {\bibfnamefont {T.}~\bibnamefont {Baehr-Jones}}, \bibinfo {author} {\bibfnamefont {M.~L.}\ \bibnamefont {Fanto}}, \bibinfo {author} {\bibfnamefont {A.~M.}\ \bibnamefont {Smith}}, \bibinfo {author} {\bibfnamefont {C.~C.}\ \bibnamefont {Tison}}, \bibinfo {author} {\bibfnamefont {P.~M.}\ \bibnamefont {Alsing}}, \emph {et~al.},\ }\bibfield  {title} {\bibinfo {title} {Linear programmable nanophotonic processors},\ }\href@noop {} {\bibfield  {journal} {\bibinfo  {journal} {Optica}\ }\textbf {\bibinfo {volume} {5}},\ \bibinfo {pages} {1623} (\bibinfo {year} {2018})}\BibitemShut {NoStop}%
\bibitem [{\citenamefont {Feldmann}\ \emph {et~al.}(2021)\citenamefont {Feldmann}, \citenamefont {Youngblood}, \citenamefont {Karpov}, \citenamefont {Gehring}, \citenamefont {Li}, \citenamefont {Stappers}, \citenamefont {Le~Gallo}, \citenamefont {Fu}, \citenamefont {Lukashchuk}, \citenamefont {Raja} \emph {et~al.}}]{feldmann2021parallel}%
  \BibitemOpen
  \bibfield  {author} {\bibinfo {author} {\bibfnamefont {J.}~\bibnamefont {Feldmann}}, \bibinfo {author} {\bibfnamefont {N.}~\bibnamefont {Youngblood}}, \bibinfo {author} {\bibfnamefont {M.}~\bibnamefont {Karpov}}, \bibinfo {author} {\bibfnamefont {H.}~\bibnamefont {Gehring}}, \bibinfo {author} {\bibfnamefont {X.}~\bibnamefont {Li}}, \bibinfo {author} {\bibfnamefont {M.}~\bibnamefont {Stappers}}, \bibinfo {author} {\bibfnamefont {M.}~\bibnamefont {Le~Gallo}}, \bibinfo {author} {\bibfnamefont {X.}~\bibnamefont {Fu}}, \bibinfo {author} {\bibfnamefont {A.}~\bibnamefont {Lukashchuk}}, \bibinfo {author} {\bibfnamefont {A.~S.}\ \bibnamefont {Raja}}, \emph {et~al.},\ }\bibfield  {title} {\bibinfo {title} {Parallel convolutional processing using an integrated photonic tensor core},\ }\href@noop {} {\bibfield  {journal} {\bibinfo  {journal} {Nature}\ }\textbf {\bibinfo {volume} {589}},\ \bibinfo {pages} {52} (\bibinfo {year} {2021})}\BibitemShut {NoStop}%
\bibitem [{\citenamefont {Wang}\ \emph {et~al.}(2023{\natexlab{b}})\citenamefont {Wang}, \citenamefont {Sohoni}, \citenamefont {Wright}, \citenamefont {Stein}, \citenamefont {Ma}, \citenamefont {Onodera}, \citenamefont {Anderson},\ and\ \citenamefont {McMahon}}]{wang2023image}%
  \BibitemOpen
  \bibfield  {author} {\bibinfo {author} {\bibfnamefont {T.}~\bibnamefont {Wang}}, \bibinfo {author} {\bibfnamefont {M.~M.}\ \bibnamefont {Sohoni}}, \bibinfo {author} {\bibfnamefont {L.~G.}\ \bibnamefont {Wright}}, \bibinfo {author} {\bibfnamefont {M.~M.}\ \bibnamefont {Stein}}, \bibinfo {author} {\bibfnamefont {S.-Y.}\ \bibnamefont {Ma}}, \bibinfo {author} {\bibfnamefont {T.}~\bibnamefont {Onodera}}, \bibinfo {author} {\bibfnamefont {M.~G.}\ \bibnamefont {Anderson}},\ and\ \bibinfo {author} {\bibfnamefont {P.~L.}\ \bibnamefont {McMahon}},\ }\bibfield  {title} {\bibinfo {title} {Image sensing with multilayer nonlinear optical neural networks},\ }\href@noop {} {\bibfield  {journal} {\bibinfo  {journal} {Nature Photonics}\ }\textbf {\bibinfo {volume} {17}},\ \bibinfo {pages} {408} (\bibinfo {year} {2023}{\natexlab{b}})}\BibitemShut {NoStop}%
\bibitem [{\citenamefont {Hua}\ \emph {et~al.}(2025)\citenamefont {Hua}, \citenamefont {Divita}, \citenamefont {Yu}, \citenamefont {Peng}, \citenamefont {Roques-Carmes}, \citenamefont {Su}, \citenamefont {Chen}, \citenamefont {Bai}, \citenamefont {Zou}, \citenamefont {Zhu} \emph {et~al.}}]{hua2025integrated}%
  \BibitemOpen
  \bibfield  {author} {\bibinfo {author} {\bibfnamefont {S.}~\bibnamefont {Hua}}, \bibinfo {author} {\bibfnamefont {E.}~\bibnamefont {Divita}}, \bibinfo {author} {\bibfnamefont {S.}~\bibnamefont {Yu}}, \bibinfo {author} {\bibfnamefont {B.}~\bibnamefont {Peng}}, \bibinfo {author} {\bibfnamefont {C.}~\bibnamefont {Roques-Carmes}}, \bibinfo {author} {\bibfnamefont {Z.}~\bibnamefont {Su}}, \bibinfo {author} {\bibfnamefont {Z.}~\bibnamefont {Chen}}, \bibinfo {author} {\bibfnamefont {Y.}~\bibnamefont {Bai}}, \bibinfo {author} {\bibfnamefont {J.}~\bibnamefont {Zou}}, \bibinfo {author} {\bibfnamefont {Y.}~\bibnamefont {Zhu}}, \emph {et~al.},\ }\bibfield  {title} {\bibinfo {title} {An integrated large-scale photonic accelerator with ultralow latency},\ }\href@noop {} {\bibfield  {journal} {\bibinfo  {journal} {Nature}\ }\textbf {\bibinfo {volume} {640}},\ \bibinfo {pages} {361} (\bibinfo {year} {2025})}\BibitemShut {NoStop}%
\bibitem [{\citenamefont {Ma}\ \emph {et~al.}(2026)\citenamefont {Ma}, \citenamefont {Laydevant}, \citenamefont {Sohoni}, \citenamefont {Wright}, \citenamefont {Wang},\ and\ \citenamefont {McMahon}}]{ma2026machine}%
  \BibitemOpen
  \bibfield  {author} {\bibinfo {author} {\bibfnamefont {S.-Y.}\ \bibnamefont {Ma}}, \bibinfo {author} {\bibfnamefont {J.}~\bibnamefont {Laydevant}}, \bibinfo {author} {\bibfnamefont {M.~M.}\ \bibnamefont {Sohoni}}, \bibinfo {author} {\bibfnamefont {L.~G.}\ \bibnamefont {Wright}}, \bibinfo {author} {\bibfnamefont {T.}~\bibnamefont {Wang}},\ and\ \bibinfo {author} {\bibfnamefont {P.~L.}\ \bibnamefont {McMahon}},\ }\bibfield  {title} {\bibinfo {title} {Machine vision with small numbers of detected photons per inference},\ }\href {https://arxiv.org/abs/2603.23974} {\bibfield  {journal} {\bibinfo  {journal} {arXiv preprint arXiv:2603.23974}\ } (\bibinfo {year} {2026})}\BibitemShut {NoStop}%
\bibitem [{\citenamefont {Choi}\ \emph {et~al.}(2024)\citenamefont {Choi}, \citenamefont {Salamin}, \citenamefont {Roques-Carmes}, \citenamefont {Dangovski}, \citenamefont {Luo}, \citenamefont {Chen}, \citenamefont {Horodynski}, \citenamefont {Sloan}, \citenamefont {Uddin},\ and\ \citenamefont {Solja{\v{c}}i{\'c}}}]{choi2024photonic}%
  \BibitemOpen
  \bibfield  {author} {\bibinfo {author} {\bibfnamefont {S.}~\bibnamefont {Choi}}, \bibinfo {author} {\bibfnamefont {Y.}~\bibnamefont {Salamin}}, \bibinfo {author} {\bibfnamefont {C.}~\bibnamefont {Roques-Carmes}}, \bibinfo {author} {\bibfnamefont {R.}~\bibnamefont {Dangovski}}, \bibinfo {author} {\bibfnamefont {D.}~\bibnamefont {Luo}}, \bibinfo {author} {\bibfnamefont {Z.}~\bibnamefont {Chen}}, \bibinfo {author} {\bibfnamefont {M.}~\bibnamefont {Horodynski}}, \bibinfo {author} {\bibfnamefont {J.}~\bibnamefont {Sloan}}, \bibinfo {author} {\bibfnamefont {S.~Z.}\ \bibnamefont {Uddin}},\ and\ \bibinfo {author} {\bibfnamefont {M.}~\bibnamefont {Solja{\v{c}}i{\'c}}},\ }\bibfield  {title} {\bibinfo {title} {Photonic probabilistic machine learning using quantum vacuum noise},\ }\href@noop {} {\bibfield  {journal} {\bibinfo  {journal} {Nature Communications}\ }\textbf {\bibinfo {volume} {15}},\ \bibinfo {pages} {7760} (\bibinfo {year} {2024})}\BibitemShut {NoStop}%
\bibitem [{\citenamefont {Br{\"u}ckerhoff-Pl{\"u}ckelmann}\ \emph {et~al.}(2024)\citenamefont {Br{\"u}ckerhoff-Pl{\"u}ckelmann}, \citenamefont {Borras}, \citenamefont {Klein}, \citenamefont {Varri}, \citenamefont {Becker}, \citenamefont {Dijkstra}, \citenamefont {Br{\"u}ckerhoff}, \citenamefont {Wright}, \citenamefont {Salinga}, \citenamefont {Bhaskaran} \emph {et~al.}}]{bruckerhoff2024probabilistic}%
  \BibitemOpen
  \bibfield  {author} {\bibinfo {author} {\bibfnamefont {F.}~\bibnamefont {Br{\"u}ckerhoff-Pl{\"u}ckelmann}}, \bibinfo {author} {\bibfnamefont {H.}~\bibnamefont {Borras}}, \bibinfo {author} {\bibfnamefont {B.}~\bibnamefont {Klein}}, \bibinfo {author} {\bibfnamefont {A.}~\bibnamefont {Varri}}, \bibinfo {author} {\bibfnamefont {M.}~\bibnamefont {Becker}}, \bibinfo {author} {\bibfnamefont {J.}~\bibnamefont {Dijkstra}}, \bibinfo {author} {\bibfnamefont {M.}~\bibnamefont {Br{\"u}ckerhoff}}, \bibinfo {author} {\bibfnamefont {C.~D.}\ \bibnamefont {Wright}}, \bibinfo {author} {\bibfnamefont {M.}~\bibnamefont {Salinga}}, \bibinfo {author} {\bibfnamefont {H.}~\bibnamefont {Bhaskaran}}, \emph {et~al.},\ }\bibfield  {title} {\bibinfo {title} {Probabilistic photonic computing with chaotic light},\ }\href@noop {} {\bibfield  {journal} {\bibinfo  {journal} {Nature Communications}\ }\textbf {\bibinfo {volume} {15}},\ \bibinfo {pages} {10445} (\bibinfo {year} {2024})}\BibitemShut {NoStop}%
\bibitem [{\citenamefont {Wu}\ \emph {et~al.}(2022)\citenamefont {Wu}, \citenamefont {Yang}, \citenamefont {Yu}, \citenamefont {Peng}, \citenamefont {Takeuchi}, \citenamefont {Chen},\ and\ \citenamefont {Li}}]{wu2022harnessing}%
  \BibitemOpen
  \bibfield  {author} {\bibinfo {author} {\bibfnamefont {C.}~\bibnamefont {Wu}}, \bibinfo {author} {\bibfnamefont {X.}~\bibnamefont {Yang}}, \bibinfo {author} {\bibfnamefont {H.}~\bibnamefont {Yu}}, \bibinfo {author} {\bibfnamefont {R.}~\bibnamefont {Peng}}, \bibinfo {author} {\bibfnamefont {I.}~\bibnamefont {Takeuchi}}, \bibinfo {author} {\bibfnamefont {Y.}~\bibnamefont {Chen}},\ and\ \bibinfo {author} {\bibfnamefont {M.}~\bibnamefont {Li}},\ }\bibfield  {title} {\bibinfo {title} {Harnessing optoelectronic noises in a photonic generative network},\ }\href@noop {} {\bibfield  {journal} {\bibinfo  {journal} {Science advances}\ }\textbf {\bibinfo {volume} {8}},\ \bibinfo {pages} {eabm2956} (\bibinfo {year} {2022})}\BibitemShut {NoStop}%
\bibitem [{\citenamefont {Zhan}\ \emph {et~al.}(2024)\citenamefont {Zhan}, \citenamefont {Wang}, \citenamefont {Liu},\ and\ \citenamefont {Fu}}]{zhan2024photonic}%
  \BibitemOpen
  \bibfield  {author} {\bibinfo {author} {\bibfnamefont {Z.}~\bibnamefont {Zhan}}, \bibinfo {author} {\bibfnamefont {H.}~\bibnamefont {Wang}}, \bibinfo {author} {\bibfnamefont {Q.}~\bibnamefont {Liu}},\ and\ \bibinfo {author} {\bibfnamefont {X.}~\bibnamefont {Fu}},\ }\bibfield  {title} {\bibinfo {title} {Photonic diffractive generators through sampling noises from scattering media},\ }\href@noop {} {\bibfield  {journal} {\bibinfo  {journal} {Nature Communications}\ }\textbf {\bibinfo {volume} {15}},\ \bibinfo {pages} {10643} (\bibinfo {year} {2024})}\BibitemShut {NoStop}%
\bibitem [{\citenamefont {Chen}\ \emph {et~al.}(2025)\citenamefont {Chen}, \citenamefont {Li}, \citenamefont {Wang}, \citenamefont {Chen},\ and\ \citenamefont {Ozcan}}]{chen2025optical}%
  \BibitemOpen
  \bibfield  {author} {\bibinfo {author} {\bibfnamefont {S.}~\bibnamefont {Chen}}, \bibinfo {author} {\bibfnamefont {Y.}~\bibnamefont {Li}}, \bibinfo {author} {\bibfnamefont {Y.}~\bibnamefont {Wang}}, \bibinfo {author} {\bibfnamefont {H.}~\bibnamefont {Chen}},\ and\ \bibinfo {author} {\bibfnamefont {A.}~\bibnamefont {Ozcan}},\ }\bibfield  {title} {\bibinfo {title} {Optical generative models},\ }\href@noop {} {\bibfield  {journal} {\bibinfo  {journal} {Nature}\ }\textbf {\bibinfo {volume} {644}},\ \bibinfo {pages} {903} (\bibinfo {year} {2025})}\BibitemShut {NoStop}%
\bibitem [{\citenamefont {Yanagimoto}\ \emph {et~al.}(2026)\citenamefont {Yanagimoto}, \citenamefont {Ash}, \citenamefont {Sohoni}, \citenamefont {Stein}, \citenamefont {Zhao}, \citenamefont {Presutti}, \citenamefont {Jankowski}, \citenamefont {Wright}, \citenamefont {Onodera},\ and\ \citenamefont {McMahon}}]{yanagimoto2026programmable}%
  \BibitemOpen
  \bibfield  {author} {\bibinfo {author} {\bibfnamefont {R.}~\bibnamefont {Yanagimoto}}, \bibinfo {author} {\bibfnamefont {B.~A.}\ \bibnamefont {Ash}}, \bibinfo {author} {\bibfnamefont {M.~M.}\ \bibnamefont {Sohoni}}, \bibinfo {author} {\bibfnamefont {M.~M.}\ \bibnamefont {Stein}}, \bibinfo {author} {\bibfnamefont {Y.}~\bibnamefont {Zhao}}, \bibinfo {author} {\bibfnamefont {F.}~\bibnamefont {Presutti}}, \bibinfo {author} {\bibfnamefont {M.}~\bibnamefont {Jankowski}}, \bibinfo {author} {\bibfnamefont {L.~G.}\ \bibnamefont {Wright}}, \bibinfo {author} {\bibfnamefont {T.}~\bibnamefont {Onodera}},\ and\ \bibinfo {author} {\bibfnamefont {P.~L.}\ \bibnamefont {McMahon}},\ }\bibfield  {title} {\bibinfo {title} {Programmable on-chip nonlinear photonics},\ }\href@noop {} {\bibfield  {journal} {\bibinfo  {journal} {Nature}\ }\textbf {\bibinfo {volume} {649}},\ \bibinfo {pages} {330} (\bibinfo {year} {2026})}\BibitemShut {NoStop}%
\bibitem [{\citenamefont {Deng}\ \emph {et~al.}(2010)\citenamefont {Deng}, \citenamefont {Haug},\ and\ \citenamefont {Yamamoto}}]{deng2010exciton}%
  \BibitemOpen
  \bibfield  {author} {\bibinfo {author} {\bibfnamefont {H.}~\bibnamefont {Deng}}, \bibinfo {author} {\bibfnamefont {H.}~\bibnamefont {Haug}},\ and\ \bibinfo {author} {\bibfnamefont {Y.}~\bibnamefont {Yamamoto}},\ }\bibfield  {title} {\bibinfo {title} {Exciton-polariton bose-einstein condensation},\ }\href@noop {} {\bibfield  {journal} {\bibinfo  {journal} {Reviews of modern physics}\ }\textbf {\bibinfo {volume} {82}},\ \bibinfo {pages} {1489} (\bibinfo {year} {2010})}\BibitemShut {NoStop}%
\bibitem [{\citenamefont {Basov}\ \emph {et~al.}(2025)\citenamefont {Basov}, \citenamefont {Asenjo-Garcia}, \citenamefont {Schuck}, \citenamefont {Zhu}, \citenamefont {Rubio}, \citenamefont {Cavalleri}, \citenamefont {Delor}, \citenamefont {Fogler},\ and\ \citenamefont {Liu}}]{basov2025polaritonic}%
  \BibitemOpen
  \bibfield  {author} {\bibinfo {author} {\bibfnamefont {D.}~\bibnamefont {Basov}}, \bibinfo {author} {\bibfnamefont {A.}~\bibnamefont {Asenjo-Garcia}}, \bibinfo {author} {\bibfnamefont {P.~J.}\ \bibnamefont {Schuck}}, \bibinfo {author} {\bibfnamefont {X.}~\bibnamefont {Zhu}}, \bibinfo {author} {\bibfnamefont {A.}~\bibnamefont {Rubio}}, \bibinfo {author} {\bibfnamefont {A.}~\bibnamefont {Cavalleri}}, \bibinfo {author} {\bibfnamefont {M.}~\bibnamefont {Delor}}, \bibinfo {author} {\bibfnamefont {M.~M.}\ \bibnamefont {Fogler}},\ and\ \bibinfo {author} {\bibfnamefont {M.}~\bibnamefont {Liu}},\ }\bibfield  {title} {\bibinfo {title} {Polaritonic quantum matter},\ }\href@noop {} {\bibfield  {journal} {\bibinfo  {journal} {Nanophotonics}\ }\textbf {\bibinfo {volume} {14}},\ \bibinfo {pages} {3723} (\bibinfo {year} {2025})}\BibitemShut {NoStop}%
\bibitem [{\citenamefont {Carusotto}\ and\ \citenamefont {Ciuti}(2013)}]{carusotto2013quantum}%
  \BibitemOpen
  \bibfield  {author} {\bibinfo {author} {\bibfnamefont {I.}~\bibnamefont {Carusotto}}\ and\ \bibinfo {author} {\bibfnamefont {C.}~\bibnamefont {Ciuti}},\ }\bibfield  {title} {\bibinfo {title} {Quantum fluids of light},\ }\href@noop {} {\bibfield  {journal} {\bibinfo  {journal} {Reviews of Modern Physics}\ }\textbf {\bibinfo {volume} {85}},\ \bibinfo {pages} {299} (\bibinfo {year} {2013})}\BibitemShut {NoStop}%
\bibitem [{\citenamefont {Kasprzak}\ \emph {et~al.}(2006)\citenamefont {Kasprzak}, \citenamefont {Richard}, \citenamefont {Kundermann}, \citenamefont {Baas}, \citenamefont {Jeambrun}, \citenamefont {Keeling}, \citenamefont {Marchetti}, \citenamefont {Szyma{\'n}ska}, \citenamefont {Andr{\'e}}, \citenamefont {Staehli} \emph {et~al.}}]{kasprzak2006bose}%
  \BibitemOpen
  \bibfield  {author} {\bibinfo {author} {\bibfnamefont {J.}~\bibnamefont {Kasprzak}}, \bibinfo {author} {\bibfnamefont {M.}~\bibnamefont {Richard}}, \bibinfo {author} {\bibfnamefont {S.}~\bibnamefont {Kundermann}}, \bibinfo {author} {\bibfnamefont {A.}~\bibnamefont {Baas}}, \bibinfo {author} {\bibfnamefont {P.}~\bibnamefont {Jeambrun}}, \bibinfo {author} {\bibfnamefont {J.~M.~J.}\ \bibnamefont {Keeling}}, \bibinfo {author} {\bibfnamefont {F.~M.}\ \bibnamefont {Marchetti}}, \bibinfo {author} {\bibfnamefont {M.}~\bibnamefont {Szyma{\'n}ska}}, \bibinfo {author} {\bibfnamefont {R.}~\bibnamefont {Andr{\'e}}}, \bibinfo {author} {\bibfnamefont {J.~a.}\ \bibnamefont {Staehli}}, \emph {et~al.},\ }\bibfield  {title} {\bibinfo {title} {Bose--einstein condensation of exciton polaritons},\ }\href@noop {} {\bibfield  {journal} {\bibinfo  {journal} {Nature}\ }\textbf {\bibinfo {volume} {443}},\ \bibinfo {pages} {409} (\bibinfo {year} {2006})}\BibitemShut {NoStop}%
\bibitem [{\citenamefont {Keeling}\ and\ \citenamefont {K{\'e}na-Cohen}(2020)}]{keeling2020bose}%
  \BibitemOpen
  \bibfield  {author} {\bibinfo {author} {\bibfnamefont {J.}~\bibnamefont {Keeling}}\ and\ \bibinfo {author} {\bibfnamefont {S.}~\bibnamefont {K{\'e}na-Cohen}},\ }\bibfield  {title} {\bibinfo {title} {Bose--einstein condensation of exciton-polaritons in organic microcavities},\ }\href@noop {} {\bibfield  {journal} {\bibinfo  {journal} {Annual Review of Physical Chemistry}\ }\textbf {\bibinfo {volume} {71}},\ \bibinfo {pages} {435} (\bibinfo {year} {2020})}\BibitemShut {NoStop}%
\bibitem [{\citenamefont {Amo}\ \emph {et~al.}(2009)\citenamefont {Amo}, \citenamefont {Sanvitto}, \citenamefont {Laussy}, \citenamefont {Ballarini}, \citenamefont {Valle}, \citenamefont {Martin}, \citenamefont {Lemaitre}, \citenamefont {Bloch}, \citenamefont {Krizhanovskii}, \citenamefont {Skolnick} \emph {et~al.}}]{amo2009collective}%
  \BibitemOpen
  \bibfield  {author} {\bibinfo {author} {\bibfnamefont {A.}~\bibnamefont {Amo}}, \bibinfo {author} {\bibfnamefont {D.}~\bibnamefont {Sanvitto}}, \bibinfo {author} {\bibfnamefont {F.}~\bibnamefont {Laussy}}, \bibinfo {author} {\bibfnamefont {D.}~\bibnamefont {Ballarini}}, \bibinfo {author} {\bibfnamefont {E.~d.}\ \bibnamefont {Valle}}, \bibinfo {author} {\bibfnamefont {M.}~\bibnamefont {Martin}}, \bibinfo {author} {\bibfnamefont {A.}~\bibnamefont {Lemaitre}}, \bibinfo {author} {\bibfnamefont {J.}~\bibnamefont {Bloch}}, \bibinfo {author} {\bibfnamefont {D.}~\bibnamefont {Krizhanovskii}}, \bibinfo {author} {\bibfnamefont {M.}~\bibnamefont {Skolnick}}, \emph {et~al.},\ }\bibfield  {title} {\bibinfo {title} {Collective fluid dynamics of a polariton condensate in a semiconductor microcavity},\ }\href@noop {} {\bibfield  {journal} {\bibinfo  {journal} {Nature}\ }\textbf {\bibinfo {volume} {457}},\ \bibinfo {pages} {291} (\bibinfo {year} {2009})}\BibitemShut {NoStop}%
\bibitem [{\citenamefont {Sich}\ \emph {et~al.}(2012)\citenamefont {Sich}, \citenamefont {Krizhanovskii}, \citenamefont {Skolnick}, \citenamefont {Gorbach}, \citenamefont {Hartley}, \citenamefont {Skryabin}, \citenamefont {Cerda-M{\'e}ndez}, \citenamefont {Biermann}, \citenamefont {Hey},\ and\ \citenamefont {Santos}}]{sich2012observation}%
  \BibitemOpen
  \bibfield  {author} {\bibinfo {author} {\bibfnamefont {M.}~\bibnamefont {Sich}}, \bibinfo {author} {\bibfnamefont {D.}~\bibnamefont {Krizhanovskii}}, \bibinfo {author} {\bibfnamefont {M.}~\bibnamefont {Skolnick}}, \bibinfo {author} {\bibfnamefont {A.~V.}\ \bibnamefont {Gorbach}}, \bibinfo {author} {\bibfnamefont {R.}~\bibnamefont {Hartley}}, \bibinfo {author} {\bibfnamefont {D.~V.}\ \bibnamefont {Skryabin}}, \bibinfo {author} {\bibfnamefont {E.}~\bibnamefont {Cerda-M{\'e}ndez}}, \bibinfo {author} {\bibfnamefont {K.}~\bibnamefont {Biermann}}, \bibinfo {author} {\bibfnamefont {R.}~\bibnamefont {Hey}},\ and\ \bibinfo {author} {\bibfnamefont {P.}~\bibnamefont {Santos}},\ }\bibfield  {title} {\bibinfo {title} {Observation of bright polariton solitons in a semiconductor microcavity},\ }\href@noop {} {\bibfield  {journal} {\bibinfo  {journal} {Nature photonics}\ }\textbf {\bibinfo {volume} {6}},\ \bibinfo {pages} {50} (\bibinfo {year} {2012})}\BibitemShut {NoStop}%
\bibitem [{\citenamefont {Togan}\ \emph {et~al.}(2018)\citenamefont {Togan}, \citenamefont {Lim}, \citenamefont {Faelt}, \citenamefont {Wegscheider},\ and\ \citenamefont {Imamoglu}}]{togan2018enhanced}%
  \BibitemOpen
  \bibfield  {author} {\bibinfo {author} {\bibfnamefont {E.}~\bibnamefont {Togan}}, \bibinfo {author} {\bibfnamefont {H.-T.}\ \bibnamefont {Lim}}, \bibinfo {author} {\bibfnamefont {S.}~\bibnamefont {Faelt}}, \bibinfo {author} {\bibfnamefont {W.}~\bibnamefont {Wegscheider}},\ and\ \bibinfo {author} {\bibfnamefont {A.}~\bibnamefont {Imamoglu}},\ }\bibfield  {title} {\bibinfo {title} {Enhanced interactions between dipolar polaritons},\ }\href@noop {} {\bibfield  {journal} {\bibinfo  {journal} {Physical review letters}\ }\textbf {\bibinfo {volume} {121}},\ \bibinfo {pages} {227402} (\bibinfo {year} {2018})}\BibitemShut {NoStop}%
\bibitem [{\citenamefont {Mu{\~n}oz-Matutano}\ \emph {et~al.}(2019)\citenamefont {Mu{\~n}oz-Matutano}, \citenamefont {Wood}, \citenamefont {Johnsson}, \citenamefont {Vidal}, \citenamefont {Baragiola}, \citenamefont {Reinhard}, \citenamefont {Lema{\^\i}tre}, \citenamefont {Bloch}, \citenamefont {Amo}, \citenamefont {Nogues} \emph {et~al.}}]{munoz2019emergence}%
  \BibitemOpen
  \bibfield  {author} {\bibinfo {author} {\bibfnamefont {G.}~\bibnamefont {Mu{\~n}oz-Matutano}}, \bibinfo {author} {\bibfnamefont {A.}~\bibnamefont {Wood}}, \bibinfo {author} {\bibfnamefont {M.}~\bibnamefont {Johnsson}}, \bibinfo {author} {\bibfnamefont {X.}~\bibnamefont {Vidal}}, \bibinfo {author} {\bibfnamefont {B.~Q.}\ \bibnamefont {Baragiola}}, \bibinfo {author} {\bibfnamefont {A.}~\bibnamefont {Reinhard}}, \bibinfo {author} {\bibfnamefont {A.}~\bibnamefont {Lema{\^\i}tre}}, \bibinfo {author} {\bibfnamefont {J.}~\bibnamefont {Bloch}}, \bibinfo {author} {\bibfnamefont {A.}~\bibnamefont {Amo}}, \bibinfo {author} {\bibfnamefont {G.}~\bibnamefont {Nogues}}, \emph {et~al.},\ }\bibfield  {title} {\bibinfo {title} {Emergence of quantum correlations from interacting fibre-cavity polaritons},\ }\href@noop {} {\bibfield  {journal} {\bibinfo  {journal} {Nature materials}\ }\textbf {\bibinfo {volume} {18}},\ \bibinfo {pages} {213} (\bibinfo {year} {2019})}\BibitemShut {NoStop}%
\bibitem [{\citenamefont {Delteil}\ \emph {et~al.}(2019)\citenamefont {Delteil}, \citenamefont {Fink}, \citenamefont {Schade}, \citenamefont {H{\"o}fling}, \citenamefont {Schneider},\ and\ \citenamefont {{\.I}mamo{\u{g}}lu}}]{delteil2019towards}%
  \BibitemOpen
  \bibfield  {author} {\bibinfo {author} {\bibfnamefont {A.}~\bibnamefont {Delteil}}, \bibinfo {author} {\bibfnamefont {T.}~\bibnamefont {Fink}}, \bibinfo {author} {\bibfnamefont {A.}~\bibnamefont {Schade}}, \bibinfo {author} {\bibfnamefont {S.}~\bibnamefont {H{\"o}fling}}, \bibinfo {author} {\bibfnamefont {C.}~\bibnamefont {Schneider}},\ and\ \bibinfo {author} {\bibfnamefont {A.}~\bibnamefont {{\.I}mamo{\u{g}}lu}},\ }\bibfield  {title} {\bibinfo {title} {Towards polariton blockade of confined exciton--polaritons},\ }\href@noop {} {\bibfield  {journal} {\bibinfo  {journal} {Nature materials}\ }\textbf {\bibinfo {volume} {18}},\ \bibinfo {pages} {219} (\bibinfo {year} {2019})}\BibitemShut {NoStop}%
\bibitem [{\citenamefont {Kuriakose}\ \emph {et~al.}(2022)\citenamefont {Kuriakose}, \citenamefont {Walker}, \citenamefont {Dowling}, \citenamefont {Kyriienko}, \citenamefont {Shelykh}, \citenamefont {St-Jean}, \citenamefont {Zambon}, \citenamefont {Lema{\^\i}tre}, \citenamefont {Sagnes}, \citenamefont {Legratiet} \emph {et~al.}}]{kuriakose2022few}%
  \BibitemOpen
  \bibfield  {author} {\bibinfo {author} {\bibfnamefont {T.}~\bibnamefont {Kuriakose}}, \bibinfo {author} {\bibfnamefont {P.~M.}\ \bibnamefont {Walker}}, \bibinfo {author} {\bibfnamefont {T.}~\bibnamefont {Dowling}}, \bibinfo {author} {\bibfnamefont {O.}~\bibnamefont {Kyriienko}}, \bibinfo {author} {\bibfnamefont {I.~A.}\ \bibnamefont {Shelykh}}, \bibinfo {author} {\bibfnamefont {P.}~\bibnamefont {St-Jean}}, \bibinfo {author} {\bibfnamefont {N.~C.}\ \bibnamefont {Zambon}}, \bibinfo {author} {\bibfnamefont {A.}~\bibnamefont {Lema{\^\i}tre}}, \bibinfo {author} {\bibfnamefont {I.}~\bibnamefont {Sagnes}}, \bibinfo {author} {\bibfnamefont {L.}~\bibnamefont {Legratiet}}, \emph {et~al.},\ }\bibfield  {title} {\bibinfo {title} {Few-photon all-optical phase rotation in a quantum-well micropillar cavity},\ }\href@noop {} {\bibfield  {journal} {\bibinfo  {journal} {Nature Photonics}\ }\textbf {\bibinfo {volume} {16}},\ \bibinfo {pages} {566} (\bibinfo {year} {2022})}\BibitemShut {NoStop}%
\bibitem [{\citenamefont {Yagafarov}\ \emph {et~al.}(2020)\citenamefont {Yagafarov}, \citenamefont {Sannikov}, \citenamefont {Zasedatelev}, \citenamefont {Georgiou}, \citenamefont {Baranikov}, \citenamefont {Kyriienko}, \citenamefont {Shelykh}, \citenamefont {Gai}, \citenamefont {Shen}, \citenamefont {Lidzey} \emph {et~al.}}]{yagafarov2020mechanisms}%
  \BibitemOpen
  \bibfield  {author} {\bibinfo {author} {\bibfnamefont {T.}~\bibnamefont {Yagafarov}}, \bibinfo {author} {\bibfnamefont {D.}~\bibnamefont {Sannikov}}, \bibinfo {author} {\bibfnamefont {A.}~\bibnamefont {Zasedatelev}}, \bibinfo {author} {\bibfnamefont {K.}~\bibnamefont {Georgiou}}, \bibinfo {author} {\bibfnamefont {A.}~\bibnamefont {Baranikov}}, \bibinfo {author} {\bibfnamefont {O.}~\bibnamefont {Kyriienko}}, \bibinfo {author} {\bibfnamefont {I.}~\bibnamefont {Shelykh}}, \bibinfo {author} {\bibfnamefont {L.}~\bibnamefont {Gai}}, \bibinfo {author} {\bibfnamefont {Z.}~\bibnamefont {Shen}}, \bibinfo {author} {\bibfnamefont {D.}~\bibnamefont {Lidzey}}, \emph {et~al.},\ }\bibfield  {title} {\bibinfo {title} {Mechanisms of blueshifts in organic polariton condensates},\ }\href@noop {} {\bibfield  {journal} {\bibinfo  {journal} {Communications Physics}\ }\textbf {\bibinfo {volume} {3}},\ \bibinfo {pages} {18} (\bibinfo {year} {2020})}\BibitemShut {NoStop}%
\bibitem [{\citenamefont {Kyriienko}\ \emph {et~al.}(2020)\citenamefont {Kyriienko}, \citenamefont {Krizhanovskii},\ and\ \citenamefont {Shelykh}}]{kyriienko2020nonlinear}%
  \BibitemOpen
  \bibfield  {author} {\bibinfo {author} {\bibfnamefont {O.}~\bibnamefont {Kyriienko}}, \bibinfo {author} {\bibfnamefont {D.}~\bibnamefont {Krizhanovskii}},\ and\ \bibinfo {author} {\bibfnamefont {I.}~\bibnamefont {Shelykh}},\ }\bibfield  {title} {\bibinfo {title} {Nonlinear quantum optics with trion polaritons in 2d monolayers: conventional and unconventional photon blockade},\ }\href@noop {} {\bibfield  {journal} {\bibinfo  {journal} {Physical Review Letters}\ }\textbf {\bibinfo {volume} {125}},\ \bibinfo {pages} {197402} (\bibinfo {year} {2020})}\BibitemShut {NoStop}%
\bibitem [{\citenamefont {Song}\ \emph {et~al.}(2024)\citenamefont {Song}, \citenamefont {Chiavazzo},\ and\ \citenamefont {Kyriienko}}]{song2024microscopic}%
  \BibitemOpen
  \bibfield  {author} {\bibinfo {author} {\bibfnamefont {K.~W.}\ \bibnamefont {Song}}, \bibinfo {author} {\bibfnamefont {S.}~\bibnamefont {Chiavazzo}},\ and\ \bibinfo {author} {\bibfnamefont {O.}~\bibnamefont {Kyriienko}},\ }\bibfield  {title} {\bibinfo {title} {Microscopic theory of nonlinear phase space filling in polaritonic lattices},\ }\href@noop {} {\bibfield  {journal} {\bibinfo  {journal} {Physical Review Research}\ }\textbf {\bibinfo {volume} {6}},\ \bibinfo {pages} {023033} (\bibinfo {year} {2024})}\BibitemShut {NoStop}%
\bibitem [{\citenamefont {Struve}\ \emph {et~al.}(2026)\citenamefont {Struve}, \citenamefont {Bennenhei}, \citenamefont {Pashaei~Adl}, \citenamefont {Song}, \citenamefont {Shan}, \citenamefont {Matukhno}, \citenamefont {Drawer}, \citenamefont {Stephan}, \citenamefont {Eilenberger}, \citenamefont {Jasti} \emph {et~al.}}]{struve2026room}%
  \BibitemOpen
  \bibfield  {author} {\bibinfo {author} {\bibfnamefont {M.}~\bibnamefont {Struve}}, \bibinfo {author} {\bibfnamefont {C.}~\bibnamefont {Bennenhei}}, \bibinfo {author} {\bibfnamefont {H.}~\bibnamefont {Pashaei~Adl}}, \bibinfo {author} {\bibfnamefont {K.~W.}\ \bibnamefont {Song}}, \bibinfo {author} {\bibfnamefont {H.}~\bibnamefont {Shan}}, \bibinfo {author} {\bibfnamefont {N.}~\bibnamefont {Matukhno}}, \bibinfo {author} {\bibfnamefont {J.-C.}\ \bibnamefont {Drawer}}, \bibinfo {author} {\bibfnamefont {S.}~\bibnamefont {Stephan}}, \bibinfo {author} {\bibfnamefont {F.}~\bibnamefont {Eilenberger}}, \bibinfo {author} {\bibfnamefont {N.~P.}\ \bibnamefont {Jasti}}, \emph {et~al.},\ }\bibfield  {title} {\bibinfo {title} {Room-temperature polariton condensate in a quasi-2d hybrid perovskite},\ }\href@noop {} {\bibfield  {journal} {\bibinfo  {journal} {Nature Communications}\ } (\bibinfo {year} {2026})}\BibitemShut {NoStop}%
\bibitem [{\citenamefont {K{\'e}na-Cohen}\ and\ \citenamefont {Forrest}(2010)}]{kena2010room}%
  \BibitemOpen
  \bibfield  {author} {\bibinfo {author} {\bibfnamefont {S.}~\bibnamefont {K{\'e}na-Cohen}}\ and\ \bibinfo {author} {\bibfnamefont {S.}~\bibnamefont {Forrest}},\ }\bibfield  {title} {\bibinfo {title} {Room-temperature polariton lasing in an organic single-crystal microcavity},\ }\href@noop {} {\bibfield  {journal} {\bibinfo  {journal} {Nature Photonics}\ }\textbf {\bibinfo {volume} {4}},\ \bibinfo {pages} {371} (\bibinfo {year} {2010})}\BibitemShut {NoStop}%
\bibitem [{\citenamefont {Plumhof}\ \emph {et~al.}(2014)\citenamefont {Plumhof}, \citenamefont {St{\"o}ferle}, \citenamefont {Mai}, \citenamefont {Scherf},\ and\ \citenamefont {Mahrt}}]{plumhof2014room}%
  \BibitemOpen
  \bibfield  {author} {\bibinfo {author} {\bibfnamefont {J.~D.}\ \bibnamefont {Plumhof}}, \bibinfo {author} {\bibfnamefont {T.}~\bibnamefont {St{\"o}ferle}}, \bibinfo {author} {\bibfnamefont {L.}~\bibnamefont {Mai}}, \bibinfo {author} {\bibfnamefont {U.}~\bibnamefont {Scherf}},\ and\ \bibinfo {author} {\bibfnamefont {R.~F.}\ \bibnamefont {Mahrt}},\ }\bibfield  {title} {\bibinfo {title} {Room-temperature bose--einstein condensation of cavity exciton--polaritons in a polymer},\ }\href@noop {} {\bibfield  {journal} {\bibinfo  {journal} {Nature materials}\ }\textbf {\bibinfo {volume} {13}},\ \bibinfo {pages} {247} (\bibinfo {year} {2014})}\BibitemShut {NoStop}%
\bibitem [{\citenamefont {Zhao}\ \emph {et~al.}(2022)\citenamefont {Zhao}, \citenamefont {Fieramosca}, \citenamefont {Bao}, \citenamefont {Du}, \citenamefont {Dini}, \citenamefont {Su}, \citenamefont {Feng}, \citenamefont {Luo}, \citenamefont {Sanvitto}, \citenamefont {Liew} \emph {et~al.}}]{zhao2022nonlinear}%
  \BibitemOpen
  \bibfield  {author} {\bibinfo {author} {\bibfnamefont {J.}~\bibnamefont {Zhao}}, \bibinfo {author} {\bibfnamefont {A.}~\bibnamefont {Fieramosca}}, \bibinfo {author} {\bibfnamefont {R.}~\bibnamefont {Bao}}, \bibinfo {author} {\bibfnamefont {W.}~\bibnamefont {Du}}, \bibinfo {author} {\bibfnamefont {K.}~\bibnamefont {Dini}}, \bibinfo {author} {\bibfnamefont {R.}~\bibnamefont {Su}}, \bibinfo {author} {\bibfnamefont {J.}~\bibnamefont {Feng}}, \bibinfo {author} {\bibfnamefont {Y.}~\bibnamefont {Luo}}, \bibinfo {author} {\bibfnamefont {D.}~\bibnamefont {Sanvitto}}, \bibinfo {author} {\bibfnamefont {T.~C.}\ \bibnamefont {Liew}}, \emph {et~al.},\ }\bibfield  {title} {\bibinfo {title} {Nonlinear polariton parametric emission in an atomically thin semiconductor based microcavity},\ }\href@noop {} {\bibfield  {journal} {\bibinfo  {journal} {Nature Nanotechnology}\ }\textbf {\bibinfo {volume} {17}},\ \bibinfo {pages} {396} (\bibinfo {year} {2022})}\BibitemShut {NoStop}%
\bibitem [{\citenamefont {Deshmukh}\ \emph {et~al.}(2024)\citenamefont {Deshmukh}, \citenamefont {Satapathy}, \citenamefont {Michail}, \citenamefont {Olsson}, \citenamefont {Bushati}, \citenamefont {Yadav}, \citenamefont {Khatoniar}, \citenamefont {Chen}, \citenamefont {John}, \citenamefont {Laursen} \emph {et~al.}}]{deshmukh2024plug}%
  \BibitemOpen
  \bibfield  {author} {\bibinfo {author} {\bibfnamefont {P.}~\bibnamefont {Deshmukh}}, \bibinfo {author} {\bibfnamefont {S.}~\bibnamefont {Satapathy}}, \bibinfo {author} {\bibfnamefont {E.}~\bibnamefont {Michail}}, \bibinfo {author} {\bibfnamefont {A.~H.}\ \bibnamefont {Olsson}}, \bibinfo {author} {\bibfnamefont {R.}~\bibnamefont {Bushati}}, \bibinfo {author} {\bibfnamefont {R.~K.}\ \bibnamefont {Yadav}}, \bibinfo {author} {\bibfnamefont {M.}~\bibnamefont {Khatoniar}}, \bibinfo {author} {\bibfnamefont {J.}~\bibnamefont {Chen}}, \bibinfo {author} {\bibfnamefont {G.}~\bibnamefont {John}}, \bibinfo {author} {\bibfnamefont {B.~W.}\ \bibnamefont {Laursen}}, \emph {et~al.},\ }\bibfield  {title} {\bibinfo {title} {Plug-and-play molecular approach for room temperature polariton condensation},\ }\href@noop {} {\bibfield  {journal} {\bibinfo  {journal} {Acs Photonics}\ }\textbf {\bibinfo {volume} {11}},\ \bibinfo {pages} {348} (\bibinfo {year} {2024})}\BibitemShut {NoStop}%
\bibitem [{\citenamefont {Georgakilas}\ \emph {et~al.}(2025)\citenamefont {Georgakilas}, \citenamefont {Tiede}, \citenamefont {Urbonas}, \citenamefont {Mirek}, \citenamefont {Bujalance}, \citenamefont {Cali{\`o}}, \citenamefont {Oddi}, \citenamefont {Tao}, \citenamefont {Dirin}, \citenamefont {Rain{\`o}} \emph {et~al.}}]{georgakilas2025room}%
  \BibitemOpen
  \bibfield  {author} {\bibinfo {author} {\bibfnamefont {I.}~\bibnamefont {Georgakilas}}, \bibinfo {author} {\bibfnamefont {D.}~\bibnamefont {Tiede}}, \bibinfo {author} {\bibfnamefont {D.}~\bibnamefont {Urbonas}}, \bibinfo {author} {\bibfnamefont {R.}~\bibnamefont {Mirek}}, \bibinfo {author} {\bibfnamefont {C.}~\bibnamefont {Bujalance}}, \bibinfo {author} {\bibfnamefont {L.}~\bibnamefont {Cali{\`o}}}, \bibinfo {author} {\bibfnamefont {V.}~\bibnamefont {Oddi}}, \bibinfo {author} {\bibfnamefont {R.}~\bibnamefont {Tao}}, \bibinfo {author} {\bibfnamefont {D.~N.}\ \bibnamefont {Dirin}}, \bibinfo {author} {\bibfnamefont {G.}~\bibnamefont {Rain{\`o}}}, \emph {et~al.},\ }\bibfield  {title} {\bibinfo {title} {Room-temperature cavity exciton-polariton condensation in perovskite quantum dots},\ }\href@noop {} {\bibfield  {journal} {\bibinfo  {journal} {Nature Communications}\ }\textbf {\bibinfo {volume} {16}},\ \bibinfo {pages} {5228} (\bibinfo {year} {2025})}\BibitemShut {NoStop}%
\bibitem [{\citenamefont {Dusel}\ \emph {et~al.}(2020)\citenamefont {Dusel}, \citenamefont {Betzold}, \citenamefont {Egorov}, \citenamefont {Klembt}, \citenamefont {Ohmer}, \citenamefont {Fischer}, \citenamefont {H{\"o}fling},\ and\ \citenamefont {Schneider}}]{dusel2020room}%
  \BibitemOpen
  \bibfield  {author} {\bibinfo {author} {\bibfnamefont {M.}~\bibnamefont {Dusel}}, \bibinfo {author} {\bibfnamefont {S.}~\bibnamefont {Betzold}}, \bibinfo {author} {\bibfnamefont {O.~A.}\ \bibnamefont {Egorov}}, \bibinfo {author} {\bibfnamefont {S.}~\bibnamefont {Klembt}}, \bibinfo {author} {\bibfnamefont {J.}~\bibnamefont {Ohmer}}, \bibinfo {author} {\bibfnamefont {U.}~\bibnamefont {Fischer}}, \bibinfo {author} {\bibfnamefont {S.}~\bibnamefont {H{\"o}fling}},\ and\ \bibinfo {author} {\bibfnamefont {C.}~\bibnamefont {Schneider}},\ }\bibfield  {title} {\bibinfo {title} {Room temperature organic exciton--polariton condensate in a lattice},\ }\href@noop {} {\bibfield  {journal} {\bibinfo  {journal} {Nature communications}\ }\textbf {\bibinfo {volume} {11}},\ \bibinfo {pages} {2863} (\bibinfo {year} {2020})}\BibitemShut {NoStop}%
\bibitem [{\citenamefont {Betzold}\ \emph {et~al.}(2024)\citenamefont {Betzold}, \citenamefont {D{\"u}reth}, \citenamefont {Dusel}, \citenamefont {Emmerling}, \citenamefont {Bieganowska}, \citenamefont {Ohmer}, \citenamefont {Fischer}, \citenamefont {H{\"o}fling},\ and\ \citenamefont {Klembt}}]{betzold2024dirac}%
  \BibitemOpen
  \bibfield  {author} {\bibinfo {author} {\bibfnamefont {S.}~\bibnamefont {Betzold}}, \bibinfo {author} {\bibfnamefont {J.}~\bibnamefont {D{\"u}reth}}, \bibinfo {author} {\bibfnamefont {M.}~\bibnamefont {Dusel}}, \bibinfo {author} {\bibfnamefont {M.}~\bibnamefont {Emmerling}}, \bibinfo {author} {\bibfnamefont {A.}~\bibnamefont {Bieganowska}}, \bibinfo {author} {\bibfnamefont {J.}~\bibnamefont {Ohmer}}, \bibinfo {author} {\bibfnamefont {U.}~\bibnamefont {Fischer}}, \bibinfo {author} {\bibfnamefont {S.}~\bibnamefont {H{\"o}fling}},\ and\ \bibinfo {author} {\bibfnamefont {S.}~\bibnamefont {Klembt}},\ }\bibfield  {title} {\bibinfo {title} {Dirac cones and room temperature polariton lasing evidenced in an organic honeycomb lattice},\ }\href@noop {} {\bibfield  {journal} {\bibinfo  {journal} {Advanced Science}\ }\textbf {\bibinfo {volume} {11}},\ \bibinfo {pages} {2400672} (\bibinfo {year} {2024})}\BibitemShut {NoStop}%
\bibitem [{\citenamefont {Opala}\ and\ \citenamefont {Matuszewski}(2023)}]{opala2023harnessing}%
  \BibitemOpen
  \bibfield  {author} {\bibinfo {author} {\bibfnamefont {A.}~\bibnamefont {Opala}}\ and\ \bibinfo {author} {\bibfnamefont {M.}~\bibnamefont {Matuszewski}},\ }\bibfield  {title} {\bibinfo {title} {Harnessing exciton-polaritons for digital computing, neuromorphic computing, and optimization},\ }\href@noop {} {\bibfield  {journal} {\bibinfo  {journal} {Optical Materials Express}\ }\textbf {\bibinfo {volume} {13}},\ \bibinfo {pages} {2674} (\bibinfo {year} {2023})}\BibitemShut {NoStop}%
\bibitem [{\citenamefont {Ghosh}\ \emph {et~al.}(2019)\citenamefont {Ghosh}, \citenamefont {Opala}, \citenamefont {Matuszewski}, \citenamefont {Paterek},\ and\ \citenamefont {Liew}}]{ghosh2019quantum}%
  \BibitemOpen
  \bibfield  {author} {\bibinfo {author} {\bibfnamefont {S.}~\bibnamefont {Ghosh}}, \bibinfo {author} {\bibfnamefont {A.}~\bibnamefont {Opala}}, \bibinfo {author} {\bibfnamefont {M.}~\bibnamefont {Matuszewski}}, \bibinfo {author} {\bibfnamefont {T.}~\bibnamefont {Paterek}},\ and\ \bibinfo {author} {\bibfnamefont {T.~C.}\ \bibnamefont {Liew}},\ }\bibfield  {title} {\bibinfo {title} {Quantum reservoir processing},\ }\href@noop {} {\bibfield  {journal} {\bibinfo  {journal} {npj Quantum Information}\ }\textbf {\bibinfo {volume} {5}},\ \bibinfo {pages} {35} (\bibinfo {year} {2019})}\BibitemShut {NoStop}%
\bibitem [{\citenamefont {Ghosh}\ \emph {et~al.}(2020)\citenamefont {Ghosh}, \citenamefont {Opala}, \citenamefont {Matuszewski}, \citenamefont {Paterek},\ and\ \citenamefont {Liew}}]{ghosh2020reconstructing}%
  \BibitemOpen
  \bibfield  {author} {\bibinfo {author} {\bibfnamefont {S.}~\bibnamefont {Ghosh}}, \bibinfo {author} {\bibfnamefont {A.}~\bibnamefont {Opala}}, \bibinfo {author} {\bibfnamefont {M.}~\bibnamefont {Matuszewski}}, \bibinfo {author} {\bibfnamefont {T.}~\bibnamefont {Paterek}},\ and\ \bibinfo {author} {\bibfnamefont {T.~C.}\ \bibnamefont {Liew}},\ }\bibfield  {title} {\bibinfo {title} {Reconstructing quantum states with quantum reservoir networks},\ }\href@noop {} {\bibfield  {journal} {\bibinfo  {journal} {IEEE Transactions on Neural Networks and Learning Systems}\ }\textbf {\bibinfo {volume} {32}},\ \bibinfo {pages} {3148} (\bibinfo {year} {2020})}\BibitemShut {NoStop}%
\bibitem [{\citenamefont {Ghosh}\ \emph {et~al.}(2021)\citenamefont {Ghosh}, \citenamefont {Nakajima}, \citenamefont {Krisnanda}, \citenamefont {Fujii},\ and\ \citenamefont {Liew}}]{ghosh2021quantum}%
  \BibitemOpen
  \bibfield  {author} {\bibinfo {author} {\bibfnamefont {S.}~\bibnamefont {Ghosh}}, \bibinfo {author} {\bibfnamefont {K.}~\bibnamefont {Nakajima}}, \bibinfo {author} {\bibfnamefont {T.}~\bibnamefont {Krisnanda}}, \bibinfo {author} {\bibfnamefont {K.}~\bibnamefont {Fujii}},\ and\ \bibinfo {author} {\bibfnamefont {T.~C.}\ \bibnamefont {Liew}},\ }\bibfield  {title} {\bibinfo {title} {Quantum neuromorphic computing with reservoir computing networks},\ }\href@noop {} {\bibfield  {journal} {\bibinfo  {journal} {Advanced Quantum Technologies}\ }\textbf {\bibinfo {volume} {4}},\ \bibinfo {pages} {2100053} (\bibinfo {year} {2021})}\BibitemShut {NoStop}%
\bibitem [{\citenamefont {Ballarini}\ \emph {et~al.}(2020)\citenamefont {Ballarini}, \citenamefont {Gianfrate}, \citenamefont {Panico}, \citenamefont {Opala}, \citenamefont {Ghosh}, \citenamefont {Dominici}, \citenamefont {Ardizzone}, \citenamefont {De~Giorgi}, \citenamefont {Lerario}, \citenamefont {Gigli} \emph {et~al.}}]{ballarini2020polaritonic}%
  \BibitemOpen
  \bibfield  {author} {\bibinfo {author} {\bibfnamefont {D.}~\bibnamefont {Ballarini}}, \bibinfo {author} {\bibfnamefont {A.}~\bibnamefont {Gianfrate}}, \bibinfo {author} {\bibfnamefont {R.}~\bibnamefont {Panico}}, \bibinfo {author} {\bibfnamefont {A.}~\bibnamefont {Opala}}, \bibinfo {author} {\bibfnamefont {S.}~\bibnamefont {Ghosh}}, \bibinfo {author} {\bibfnamefont {L.}~\bibnamefont {Dominici}}, \bibinfo {author} {\bibfnamefont {V.}~\bibnamefont {Ardizzone}}, \bibinfo {author} {\bibfnamefont {M.}~\bibnamefont {De~Giorgi}}, \bibinfo {author} {\bibfnamefont {G.}~\bibnamefont {Lerario}}, \bibinfo {author} {\bibfnamefont {G.}~\bibnamefont {Gigli}}, \emph {et~al.},\ }\bibfield  {title} {\bibinfo {title} {Polaritonic neuromorphic computing outperforms linear classifiers},\ }\href@noop {} {\bibfield  {journal} {\bibinfo  {journal} {Nano Letters}\ }\textbf {\bibinfo {volume} {20}},\ \bibinfo {pages} {3506} (\bibinfo {year} {2020})}\BibitemShut {NoStop}%
\bibitem [{\citenamefont {Mirek}\ \emph {et~al.}(2021)\citenamefont {Mirek}, \citenamefont {Opala}, \citenamefont {Comaron}, \citenamefont {Furman}, \citenamefont {Kr{\'o}l}, \citenamefont {Tyszka}, \citenamefont {Seredynski}, \citenamefont {Ballarini}, \citenamefont {Sanvitto}, \citenamefont {Liew} \emph {et~al.}}]{mirek2021neuromorphic}%
  \BibitemOpen
  \bibfield  {author} {\bibinfo {author} {\bibfnamefont {R.}~\bibnamefont {Mirek}}, \bibinfo {author} {\bibfnamefont {A.}~\bibnamefont {Opala}}, \bibinfo {author} {\bibfnamefont {P.}~\bibnamefont {Comaron}}, \bibinfo {author} {\bibfnamefont {M.}~\bibnamefont {Furman}}, \bibinfo {author} {\bibfnamefont {M.}~\bibnamefont {Kr{\'o}l}}, \bibinfo {author} {\bibfnamefont {K.}~\bibnamefont {Tyszka}}, \bibinfo {author} {\bibfnamefont {B.}~\bibnamefont {Seredynski}}, \bibinfo {author} {\bibfnamefont {D.}~\bibnamefont {Ballarini}}, \bibinfo {author} {\bibfnamefont {D.}~\bibnamefont {Sanvitto}}, \bibinfo {author} {\bibfnamefont {T.~C.}\ \bibnamefont {Liew}}, \emph {et~al.},\ }\bibfield  {title} {\bibinfo {title} {Neuromorphic binarized polariton networks},\ }\href@noop {} {\bibfield  {journal} {\bibinfo  {journal} {Nano letters}\ }\textbf {\bibinfo {volume} {21}},\ \bibinfo {pages} {3715} (\bibinfo {year} {2021})}\BibitemShut {NoStop}%
\bibitem [{\citenamefont {K{\k{e}}dziora}\ \emph {et~al.}(2024)\citenamefont {K{\k{e}}dziora}, \citenamefont {Opala}, \citenamefont {Mastria}, \citenamefont {De~Marco}, \citenamefont {Kr{\'o}l}, \citenamefont {{\L}empicka-Mirek}, \citenamefont {Tyszka}, \citenamefont {Ekielski}, \citenamefont {Guziewicz}, \citenamefont {Bogdanowicz} \emph {et~al.}}]{kkedziora2024predesigned}%
  \BibitemOpen
  \bibfield  {author} {\bibinfo {author} {\bibfnamefont {M.}~\bibnamefont {K{\k{e}}dziora}}, \bibinfo {author} {\bibfnamefont {A.}~\bibnamefont {Opala}}, \bibinfo {author} {\bibfnamefont {R.}~\bibnamefont {Mastria}}, \bibinfo {author} {\bibfnamefont {L.}~\bibnamefont {De~Marco}}, \bibinfo {author} {\bibfnamefont {M.}~\bibnamefont {Kr{\'o}l}}, \bibinfo {author} {\bibfnamefont {K.}~\bibnamefont {{\L}empicka-Mirek}}, \bibinfo {author} {\bibfnamefont {K.}~\bibnamefont {Tyszka}}, \bibinfo {author} {\bibfnamefont {M.}~\bibnamefont {Ekielski}}, \bibinfo {author} {\bibfnamefont {M.}~\bibnamefont {Guziewicz}}, \bibinfo {author} {\bibfnamefont {K.}~\bibnamefont {Bogdanowicz}}, \emph {et~al.},\ }\bibfield  {title} {\bibinfo {title} {Predesigned perovskite crystal waveguides for room-temperature exciton--polariton condensation and edge lasing},\ }\href@noop {} {\bibfield  {journal} {\bibinfo  {journal} {Nature Materials}\ }\textbf {\bibinfo {volume} {23}},\ \bibinfo {pages} {1515} (\bibinfo {year} {2024})}\BibitemShut
  {NoStop}%
\bibitem [{\citenamefont {Opala}\ \emph {et~al.}(2024)\citenamefont {Opala}, \citenamefont {Tyszka}, \citenamefont {K{\k{e}}dziora}, \citenamefont {Furman}, \citenamefont {Rahmani}, \citenamefont {{\'S}wierczewski}, \citenamefont {Ekielski}, \citenamefont {Szerling}, \citenamefont {Matuszewski},\ and\ \citenamefont {Pi{\k{e}}tka}}]{opala2024room}%
  \BibitemOpen
  \bibfield  {author} {\bibinfo {author} {\bibfnamefont {A.}~\bibnamefont {Opala}}, \bibinfo {author} {\bibfnamefont {K.}~\bibnamefont {Tyszka}}, \bibinfo {author} {\bibfnamefont {M.}~\bibnamefont {K{\k{e}}dziora}}, \bibinfo {author} {\bibfnamefont {M.}~\bibnamefont {Furman}}, \bibinfo {author} {\bibfnamefont {A.}~\bibnamefont {Rahmani}}, \bibinfo {author} {\bibfnamefont {S.}~\bibnamefont {{\'S}wierczewski}}, \bibinfo {author} {\bibfnamefont {M.}~\bibnamefont {Ekielski}}, \bibinfo {author} {\bibfnamefont {A.}~\bibnamefont {Szerling}}, \bibinfo {author} {\bibfnamefont {M.}~\bibnamefont {Matuszewski}},\ and\ \bibinfo {author} {\bibfnamefont {B.}~\bibnamefont {Pi{\k{e}}tka}},\ }\bibfield  {title} {\bibinfo {title} {Room temperature exciton-polariton neural network with perovskite crystal},\ }\href {https://arxiv.org/abs/2412.10865} {\bibfield  {journal} {\bibinfo  {journal} {arXiv preprint arXiv:2412.10865}\ } (\bibinfo {year} {2024})}\BibitemShut {NoStop}%
\bibitem [{\citenamefont {Zaremba}\ \emph {et~al.}(2025)\citenamefont {Zaremba}, \citenamefont {K{\k{e}}dziora}, \citenamefont {Sta{\'n}co}, \citenamefont {Piskorski}, \citenamefont {Kosiel}, \citenamefont {Szerling}, \citenamefont {Mazur}, \citenamefont {Piecek}, \citenamefont {Opala}, \citenamefont {Sigur{\dh}sson} \emph {et~al.}}]{zaremba2025optically}%
  \BibitemOpen
  \bibfield  {author} {\bibinfo {author} {\bibfnamefont {M.}~\bibnamefont {Zaremba}}, \bibinfo {author} {\bibfnamefont {M.}~\bibnamefont {K{\k{e}}dziora}}, \bibinfo {author} {\bibfnamefont {L.}~\bibnamefont {Sta{\'n}co}}, \bibinfo {author} {\bibfnamefont {K.}~\bibnamefont {Piskorski}}, \bibinfo {author} {\bibfnamefont {K.}~\bibnamefont {Kosiel}}, \bibinfo {author} {\bibfnamefont {A.}~\bibnamefont {Szerling}}, \bibinfo {author} {\bibfnamefont {R.}~\bibnamefont {Mazur}}, \bibinfo {author} {\bibfnamefont {W.}~\bibnamefont {Piecek}}, \bibinfo {author} {\bibfnamefont {A.}~\bibnamefont {Opala}}, \bibinfo {author} {\bibfnamefont {H.}~\bibnamefont {Sigur{\dh}sson}}, \emph {et~al.},\ }\bibfield  {title} {\bibinfo {title} {Optically trapped exciton-polariton condensates in a perovskite microcavity},\ }\href@noop {} {\bibfield  {journal} {\bibinfo  {journal} {Advanced Optical Materials}\ }\textbf {\bibinfo {volume} {13}},\ \bibinfo {pages} {2500304} (\bibinfo {year} {2025})}\BibitemShut {NoStop}%
\bibitem [{\citenamefont {Gan}\ \emph {et~al.}(2025)\citenamefont {Gan}, \citenamefont {Shi}, \citenamefont {Ghosh}, \citenamefont {Liu}, \citenamefont {Xu},\ and\ \citenamefont {Xiong}}]{gan2025ultrafast}%
  \BibitemOpen
  \bibfield  {author} {\bibinfo {author} {\bibfnamefont {Y.}~\bibnamefont {Gan}}, \bibinfo {author} {\bibfnamefont {Y.}~\bibnamefont {Shi}}, \bibinfo {author} {\bibfnamefont {S.}~\bibnamefont {Ghosh}}, \bibinfo {author} {\bibfnamefont {H.}~\bibnamefont {Liu}}, \bibinfo {author} {\bibfnamefont {H.}~\bibnamefont {Xu}},\ and\ \bibinfo {author} {\bibfnamefont {Q.}~\bibnamefont {Xiong}},\ }\bibfield  {title} {\bibinfo {title} {Ultrafast neuromorphic computing driven by polariton nonlinearities},\ }\href@noop {} {\bibfield  {journal} {\bibinfo  {journal} {Elight}\ }\textbf {\bibinfo {volume} {5}},\ \bibinfo {pages} {9} (\bibinfo {year} {2025})}\BibitemShut {NoStop}%
\bibitem [{\citenamefont {Wang}\ \emph {et~al.}(2025)\citenamefont {Wang}, \citenamefont {Scali},\ and\ \citenamefont {Kyriienko}}]{wang2025polaritonic}%
  \BibitemOpen
  \bibfield  {author} {\bibinfo {author} {\bibfnamefont {Y.}~\bibnamefont {Wang}}, \bibinfo {author} {\bibfnamefont {S.}~\bibnamefont {Scali}},\ and\ \bibinfo {author} {\bibfnamefont {O.}~\bibnamefont {Kyriienko}},\ }\bibfield  {title} {\bibinfo {title} {Polaritonic machine learning for graph-based data analysis},\ }\href {https://arxiv.org/abs/2507.10415} {\bibfield  {journal} {\bibinfo  {journal} {arXiv preprint arXiv:2507.10415}\ } (\bibinfo {year} {2025})}\BibitemShut {NoStop}%
\bibitem [{\citenamefont {Wang}\ and\ \citenamefont {Kyriienko}(2025)}]{wang2025photonics}%
  \BibitemOpen
  \bibfield  {author} {\bibinfo {author} {\bibfnamefont {Y.}~\bibnamefont {Wang}}\ and\ \bibinfo {author} {\bibfnamefont {O.}~\bibnamefont {Kyriienko}},\ }\bibfield  {title} {\bibinfo {title} {Photonics-enhanced graph convolutional networks},\ }\href {https://arxiv.org/abs/2512.15549} {\bibfield  {journal} {\bibinfo  {journal} {arXiv preprint arXiv:2512.15549}\ } (\bibinfo {year} {2025})}\BibitemShut {NoStop}%
\bibitem [{\citenamefont {Gulrajani}\ \emph {et~al.}(2017)\citenamefont {Gulrajani}, \citenamefont {Ahmed}, \citenamefont {Arjovsky}, \citenamefont {Dumoulin},\ and\ \citenamefont {Courville}}]{gulrajani2017improved}%
  \BibitemOpen
  \bibfield  {author} {\bibinfo {author} {\bibfnamefont {I.}~\bibnamefont {Gulrajani}}, \bibinfo {author} {\bibfnamefont {F.}~\bibnamefont {Ahmed}}, \bibinfo {author} {\bibfnamefont {M.}~\bibnamefont {Arjovsky}}, \bibinfo {author} {\bibfnamefont {V.}~\bibnamefont {Dumoulin}},\ and\ \bibinfo {author} {\bibfnamefont {A.~C.}\ \bibnamefont {Courville}},\ }\bibfield  {title} {\bibinfo {title} {Improved training of wasserstein gans},\ }\href@noop {} {\bibfield  {journal} {\bibinfo  {journal} {Advances in neural information processing systems}\ }\textbf {\bibinfo {volume} {30}} (\bibinfo {year} {2017})}\BibitemShut {NoStop}%
\bibitem [{\citenamefont {Arjovsky}\ \emph {et~al.}(2017)\citenamefont {Arjovsky}, \citenamefont {Chintala},\ and\ \citenamefont {Bottou}}]{arjovsky2017wasserstein}%
  \BibitemOpen
  \bibfield  {author} {\bibinfo {author} {\bibfnamefont {M.}~\bibnamefont {Arjovsky}}, \bibinfo {author} {\bibfnamefont {S.}~\bibnamefont {Chintala}},\ and\ \bibinfo {author} {\bibfnamefont {L.}~\bibnamefont {Bottou}},\ }\bibfield  {title} {\bibinfo {title} {Wasserstein generative adversarial networks},\ }in\ \href@noop {} {\emph {\bibinfo {booktitle} {International conference on machine learning}}}\ (\bibinfo {organization} {Pmlr},\ \bibinfo {year} {2017})\ pp.\ \bibinfo {pages} {214--223}\BibitemShut {NoStop}%
\bibitem [{\citenamefont {LeCun}\ \emph {et~al.}(1998)\citenamefont {LeCun}, \citenamefont {Bottou}, \citenamefont {Bengio},\ and\ \citenamefont {Haffner}}]{lecun1998gradient}%
  \BibitemOpen
  \bibfield  {author} {\bibinfo {author} {\bibfnamefont {Y.}~\bibnamefont {LeCun}}, \bibinfo {author} {\bibfnamefont {L.}~\bibnamefont {Bottou}}, \bibinfo {author} {\bibfnamefont {Y.}~\bibnamefont {Bengio}},\ and\ \bibinfo {author} {\bibfnamefont {P.}~\bibnamefont {Haffner}},\ }\bibfield  {title} {\bibinfo {title} {Gradient-based learning applied to document recognition},\ }\href@noop {} {\bibfield  {journal} {\bibinfo  {journal} {Proceedings of the IEEE}\ }\textbf {\bibinfo {volume} {86}},\ \bibinfo {pages} {2278} (\bibinfo {year} {1998})}\BibitemShut {NoStop}%
\bibitem [{\citenamefont {Estrecho}\ \emph {et~al.}(2019)\citenamefont {Estrecho}, \citenamefont {Gao}, \citenamefont {Bobrovska}, \citenamefont {Comber-Todd}, \citenamefont {Fraser}, \citenamefont {Steger}, \citenamefont {West}, \citenamefont {Pfeiffer}, \citenamefont {Levinsen}, \citenamefont {Parish} \emph {et~al.}}]{estrecho2019direct}%
  \BibitemOpen
  \bibfield  {author} {\bibinfo {author} {\bibfnamefont {E.}~\bibnamefont {Estrecho}}, \bibinfo {author} {\bibfnamefont {T.}~\bibnamefont {Gao}}, \bibinfo {author} {\bibfnamefont {N.}~\bibnamefont {Bobrovska}}, \bibinfo {author} {\bibfnamefont {D.}~\bibnamefont {Comber-Todd}}, \bibinfo {author} {\bibfnamefont {M.~D.}\ \bibnamefont {Fraser}}, \bibinfo {author} {\bibfnamefont {M.}~\bibnamefont {Steger}}, \bibinfo {author} {\bibfnamefont {K.}~\bibnamefont {West}}, \bibinfo {author} {\bibfnamefont {L.~N.}\ \bibnamefont {Pfeiffer}}, \bibinfo {author} {\bibfnamefont {J.}~\bibnamefont {Levinsen}}, \bibinfo {author} {\bibfnamefont {M.}~\bibnamefont {Parish}}, \emph {et~al.},\ }\bibfield  {title} {\bibinfo {title} {Direct measurement of polariton-polariton interaction strength in the thomas-fermi regime of exciton-polariton condensation},\ }\href@noop {} {\bibfield  {journal} {\bibinfo  {journal} {Physical Review B}\ }\textbf {\bibinfo {volume} {100}},\ \bibinfo {pages} {035306} (\bibinfo {year} {2019})}\BibitemShut
  {NoStop}%
\bibitem [{\citenamefont {Daskalakis}\ \emph {et~al.}(2014)\citenamefont {Daskalakis}, \citenamefont {Maier}, \citenamefont {Murray},\ and\ \citenamefont {K{\'e}na-Cohen}}]{daskalakis2014nonlinear}%
  \BibitemOpen
  \bibfield  {author} {\bibinfo {author} {\bibfnamefont {K.}~\bibnamefont {Daskalakis}}, \bibinfo {author} {\bibfnamefont {S.}~\bibnamefont {Maier}}, \bibinfo {author} {\bibfnamefont {R.}~\bibnamefont {Murray}},\ and\ \bibinfo {author} {\bibfnamefont {S.}~\bibnamefont {K{\'e}na-Cohen}},\ }\bibfield  {title} {\bibinfo {title} {Nonlinear interactions in an organic polariton condensate},\ }\href@noop {} {\bibfield  {journal} {\bibinfo  {journal} {Nature materials}\ }\textbf {\bibinfo {volume} {13}},\ \bibinfo {pages} {271} (\bibinfo {year} {2014})}\BibitemShut {NoStop}%
\bibitem [{\citenamefont {Goodfellow}(2016)}]{goodfellow2016nips}%
  \BibitemOpen
  \bibfield  {author} {\bibinfo {author} {\bibfnamefont {I.}~\bibnamefont {Goodfellow}},\ }\bibfield  {title} {\bibinfo {title} {Nips 2016 tutorial: Generative adversarial networks},\ }\href {https://arxiv.org/abs/1701.00160} {\bibfield  {journal} {\bibinfo  {journal} {arXiv preprint arXiv:1701.00160}\ } (\bibinfo {year} {2016})}\BibitemShut {NoStop}%
\bibitem [{\citenamefont {Salimans}\ \emph {et~al.}(2016)\citenamefont {Salimans}, \citenamefont {Goodfellow}, \citenamefont {Zaremba}, \citenamefont {Cheung}, \citenamefont {Radford},\ and\ \citenamefont {Chen}}]{salimans2016improved}%
  \BibitemOpen
  \bibfield  {author} {\bibinfo {author} {\bibfnamefont {T.}~\bibnamefont {Salimans}}, \bibinfo {author} {\bibfnamefont {I.}~\bibnamefont {Goodfellow}}, \bibinfo {author} {\bibfnamefont {W.}~\bibnamefont {Zaremba}}, \bibinfo {author} {\bibfnamefont {V.}~\bibnamefont {Cheung}}, \bibinfo {author} {\bibfnamefont {A.}~\bibnamefont {Radford}},\ and\ \bibinfo {author} {\bibfnamefont {X.}~\bibnamefont {Chen}},\ }\bibfield  {title} {\bibinfo {title} {Improved techniques for training gans},\ }\href@noop {} {\bibfield  {journal} {\bibinfo  {journal} {Advances in neural information processing systems}\ }\textbf {\bibinfo {volume} {29}} (\bibinfo {year} {2016})}\BibitemShut {NoStop}%
\bibitem [{\citenamefont {Bacarreza}\ \emph {et~al.}(2025)\citenamefont {Bacarreza}, \citenamefont {Farnsworth}, \citenamefont {Makarovskiy}, \citenamefont {Wallner}, \citenamefont {Hicks}, \citenamefont {Sempere-Llagostera}, \citenamefont {Price}, \citenamefont {Francis-Jones},\ and\ \citenamefont {Clements}}]{bacarreza2025quantum}%
  \BibitemOpen
  \bibfield  {author} {\bibinfo {author} {\bibfnamefont {O.}~\bibnamefont {Bacarreza}}, \bibinfo {author} {\bibfnamefont {T.}~\bibnamefont {Farnsworth}}, \bibinfo {author} {\bibfnamefont {A.}~\bibnamefont {Makarovskiy}}, \bibinfo {author} {\bibfnamefont {H.}~\bibnamefont {Wallner}}, \bibinfo {author} {\bibfnamefont {T.}~\bibnamefont {Hicks}}, \bibinfo {author} {\bibfnamefont {S.}~\bibnamefont {Sempere-Llagostera}}, \bibinfo {author} {\bibfnamefont {J.}~\bibnamefont {Price}}, \bibinfo {author} {\bibfnamefont {R.~J.}\ \bibnamefont {Francis-Jones}},\ and\ \bibinfo {author} {\bibfnamefont {W.~R.}\ \bibnamefont {Clements}},\ }\bibfield  {title} {\bibinfo {title} {Quantum latent distributions in deep generative models},\ }\href {https://arxiv.org/abs/2508.19857} {\bibfield  {journal} {\bibinfo  {journal} {arXiv preprint arXiv:2508.19857}\ } (\bibinfo {year} {2025})}\BibitemShut {NoStop}%
\end{thebibliography}

\onecolumngrid


\section*{Supplementary material}

# Supplementary Information for Generative Modelling Powered by Room-temperature Polariton Condensates

Yuan Wang[1], Marcin Muszynski[2,3], Avinash Dash[2,3], Rishabh Kaurav[2,4], Vinod M. Menon[2,3] and Oleksandr Kyriienko[1]

[1]School of Mathematical and Physical Sciences, University of Sheffield, Sheffield, S10 2TN, United Kingdom.
[2]Department of Physics, City College of New York, New York, NY 10031, United States.
[3]Department of Physics, Graduate Center of the City University of New York, New York, NY 10016, United States.
[4]Chemistry Doctoral Program, Graduate Center of the City University of New York, New York, NY 10016, United States.

## Supplementary Note 1 Architecture of generative adversarial network

In this section, we detail the architecture of a conditional generative adversarial network (GAN) used in the main text. Our GAN consists of three main components.

**Generator:** The generator $G : \mathbb{R}^{n+c} \to \mathbb{R}^n$ is implemented as a 3-layer MLP that takes as input a concatenation of the flattened input image $\mathbf{x} \in \mathbb{R}^{428\times428}$ and a one-hot encoded class label $\mathbf{y} \in \{0,1\}^2$, where $n = 428^2 = 183,184$ and $c = 2$. With hidden dimension $d_h = 128$, the generator architecture is defined as

$$\mathbf{z}_0 = [\text{flatten}(\mathbf{x});\ \mathbf{y}_{\text{one-hot}}] \in \mathbb{R}^{183,186}, \tag{S1}$$

$$\mathbf{h}_1 = \text{LeakyReLU}(0.2, \text{BN}(\mathbf{W}_1\mathbf{z}_0 + \mathbf{b}_1)), \tag{S2}$$

$$\mathbf{h}_2 = \text{LeakyReLU}(0.2, \text{BN}(\mathbf{W}_2\mathbf{h}_1 + \mathbf{b}_2)), \tag{S3}$$

$$G(\mathbf{x}, \mathbf{y}) = \text{reshape}(\tanh(\mathbf{W}_3\mathbf{h}_2 + \mathbf{b}_3), [1, 428, 428]), \tag{S4}$$

where $\mathbf{W}_1 \in \mathbb{R}^{64\times 183,186}$, $\mathbf{W}_2 \in \mathbb{R}^{128\times 64}$, $\mathbf{W}_3 \in \mathbb{R}^{183,184\times 128}$, BN denotes batch normalization, and $\text{LeakyReLU}(\alpha, x) = \max(\alpha x, x)$ with $\alpha = 0.2$ maintaining a small gradient even for negative inputs. The architecture follows a progressive expansion pattern $183, 186 \to 64 \to 128 \to 183, 184$, where the intermediate layer ($d_h/2 = 64$) provides dimensional compression before expanding to full hidden capacity ($d_h = 128$). The output is normalized to $[-1, 1]$ using the tanh activation function.

**Critic:** Following the WGAN-GP formulation, we employ a critic $D : \mathbb{R}^{n+c} \to \mathbb{R}$ instead of a traditional discriminator. The critic evaluates the Wasserstein distance between real and generated distributions without the constraints of outputting probabilities. Critically, we omit batch normalization in the critic to maintain the validity of the gradient penalty computation. The critic architecture follows a progressive dimension reduction pattern

$$\mathbf{z}_0 = [\text{flatten}(\mathbf{x});\ \mathbf{y}_{\text{one-hot}}] \in \mathbb{R}^{183,186}, \tag{S5}$$

$$\mathbf{h}_1 = \text{LeakyReLU}(0.2, \mathbf{W}_1\mathbf{z}_0 + \mathbf{b}_1), \tag{S6}$$

$$\mathbf{h}_2 = \text{LeakyReLU}(0.2, \mathbf{W}_2\mathbf{h}_1 + \mathbf{b}_2), \tag{S7}$$

$$D(\mathbf{x}, \mathbf{y}) = \mathbf{W}_3\mathbf{h}_2 + \mathbf{b}_3, \tag{S8}$$

where $\mathbf{W}_1 \in \mathbb{R}^{64\times 183,186}$, $\mathbf{W}_2 \in \mathbb{R}^{128\times 64}$, and $\mathbf{W}_3 \in \mathbb{R}^{1\times 128}$, implementing the reduction $183, 186 \to 64 \to 128 \to 1$.

**Digit preservation classifier:** To evaluate preservation of digit identity during translation, we employ a CNN-based binary classifier $C : \mathbb{R}^{428\times 428} \to \mathbb{R}^2$ that predicts the digit class (0 or 1). The classifier is pre-trained on MNIST data and frozen during GAN training, serving solely as an evaluation metric. The architecture consists of five convolutional blocks with progressive channel expansion followed by fully connected classification layers:

$$\mathbf{f}_1 = \text{MaxPool}_2(\text{ReLU}(\text{BN}(\text{Conv}^{32}_{5\times 5, s=2, p=2}(\mathbf{x})))), \quad 428\times 428 \to 107\times 107 \tag{S9}$$

$$\mathbf{f}_2 = \text{MaxPool}_2(\text{ReLU}(\text{BN}(\text{Conv}^{64}_{3\times 3, s=1, p=1}(\mathbf{f}_1)))), \quad 107\times 107 \to 53\times 53 \tag{S10}$$

$$\mathbf{f}_3 = \text{MaxPool}_2(\text{ReLU}(\text{BN}(\text{Conv}^{128}_{3\times 3, s=1, p=1}(\mathbf{f}_2)))), \quad 53\times 53 \to 26\times 26 \tag{S11}$$

$$\mathbf{f}_4 = \text{MaxPool}_2(\text{ReLU}(\text{BN}(\text{Conv}^{256}_{3\times 3, s=1, p=1}(\mathbf{f}_3)))), \quad 26\times 26 \to 13\times 13 \tag{S12}$$

$$\mathbf{f}_5 = \text{AdaptiveAvgPool}_{1\times 1}(\text{ReLU}(\text{BN}(\text{Conv}^{256}_{3\times 3, s=1, p=1}(\mathbf{f}_4)))), \quad 13\times 13 \to 1\times 1 \tag{S13}$$

$$C(\mathbf{x}) = \text{Linear}^2(\text{ReLU}(\text{Linear}^{128}(\text{flatten}(\mathbf{f}_5)))), \tag{S14}$$

where $\text{Conv}^{c}_{k\times k,s,p}$ denotes a convolutional layer with $c$ output channels, kernel size $k \times k$, stride $s$, and padding $p$.

# Supplementary Note 2 Dataset preparation

Next, we detail the procedure for preparing the datasets used in the main text. To train the conditional GAN architecture, we construct a carefully balanced dataset of binary digit images ('0' and '1') at $428 \times 428$ pixel resolution across three categories: **(i)** experimentally captured PL images of polariton condensates / laser light reflected from an ordinary mirror, **(ii)** Helvetica font digits with random perturbation injection for comparison, and **(iii)** MNIST handwritten digits as target samples for adversarial training. All datasets maintain equal representation of both digit classes. The dataset generation process begins with a set of 8 base Helvetica digit images: 4 images of digit '0' and 4 images of digit '1'. These base images differ from each other only through slight translations along the horizontal or vertical directions, providing minimal variation while maintaining digit identity. From these base images, we generate two parallel datasets using fundamentally different mechanisms for introducing variability.

**Condensate mode:** Each of the 8 base Helvetica images is encoded via SLM and projected onto the polariton condensate system. For each SLM projection, we capture 480 successive time-integrated PL images which, despite nominally identical excitation patterns, differ from one to another due to the stochastic nature of the condensate formation process and the nonlinear polariton-polariton interactions. The evolution of these PL images across successive experimental realizations introduces variability governed by coherent many-body dynamics, producing spatially-correlated intensity fluctuations that reflect the underlying nonlinear physics of the system. This process yields $8 \times 480 = 3{,}840$ PL images with physically-generated structured variability.

**Baseline mode:** To create a controlled comparison, we apply 480 different instances of random computational perturbation injection to each of the 8 base Helvetica images. Each perturbation instance is independently sampled from a Gaussian distribution, providing computationally-generated uncorrelated variability. An analogous computational perturbation, detailed in Supplementary Note 5.2.1, is applied to the bare SLM projections reflected from a mirror, corresponding to the laser-mode control experiment discussed in the Results section of the main text. This parallel process also yields $8 \times 480 = 3{,}840$ images. Both datasets contain the same number of samples with identical class balance, enabling direct comparison of how the nature of input variability affects generative model performance. The large image dimension ($428 \times 428$ pixels) is necessary to preserve fine spatial details in PL

patterns from experimental acquisition.

Once we have data generated, we proceed to data partitioning. For GAN training, each mode (condensate and baseline) uses its respective dataset of 3,840 images. This is partitioned as follows: we use 768 images (20%) for training, 768 images (20%) for validation, and 2,304 images (60%) for testing. This partitioning maintains equal representation of both digit classes in each subset. Training pairs combine input images (either condensate or Helvetica with perturbations) with corresponding MNIST handwritten digits as target references, creating a supervised translation task. The critic learns to distinguish between generated samples and real MNIST digits, while the generator learns the conditional transformation from input images to MNIST-style outputs that preserve digit identity. Importantly, the MNIST dataset used for GAN training is distinct from the MNIST dataset used for pre-training the classifier $C$. The classifier is pre-trained on a separate MNIST subset containing 6,144 binary digit images (only '0' and '1'), partitioned into 80% for training and 20% for validation. This separation ensures no data leakage between the classifier evaluation and the GAN training process. After pre-training, the classifier is frozen and serves only as an evaluation metric during GAN training, not contributing to the adversarial process.

# Supplementary Note 3 Training procedure

Following standard WGAN-GP practice, we employ a multi-step training procedure where the critic is updated $n_{\text{critic}} = 5$ times for each generator update. The classifier $C$ is pre-trained for 20 epochs on MNIST samples using cross-entropy loss before being frozen and used only for evaluation during GAN training. We use the Adam optimizer with momentum parameters $\beta_1 = 0.5$, $\beta_2 = 0.999$ for both generator and critic. To ensure consistency and facilitate fair comparison between condensate and baseline modes, we use a generator learning rate of $1 \times 10^{-4}$ and a critic learning rate of $5 \times 10^{-5}$, applied identically across both operational modes. The slower critic update rate relative to the generator promotes stable Wasserstein estimation while preventing the critic from overwhelming the generator early in training. All weights are initialized from $\mathcal{N}(0, 0.02)$, with biases initialized to zero. Training is conducted for 100 epochs with batch size 8.

# Supplementary Note 4 Evaluation metrics

We evaluate our model using four metrics: inception score (IS), structural similarity index (SSIM), digit preservation accuracy (DPA), and pixel variance.

**Inception score:** We adapt the standard IS formulation [1] for our binary task

$$\text{IS} = \exp\left(\mathbb{E}_{\mathbf{x}\sim p_G}[\text{KL}(p(\mathbf{y}|\mathbf{x})\|p(\mathbf{y}))]\right), \tag{S15}$$

where $p(\mathbf{y}|\mathbf{x}) = \text{softmax}(C(\mathbf{x}))$ is the conditional class distribution predicted by the pre-trained classifier $C$ for generated image $\mathbf{x}$, and $p(\mathbf{y}) = \frac{1}{N}\sum_{i=1}^{N} p(\mathbf{y}|\mathbf{x}_i)$ is the marginal class distribution computed over all $N$ generated samples. The KL divergence $\text{KL}(p\|q) = \sum_y p(y)\log\frac{p(y)}{q(y)}$ measures how much the conditional distribution differs from the marginal. To estimate variance, we employ a splitting strategy: $N$ generated images are divided into $K$ splits (here, $K = 10$), and IS is computed independently for each split using the global marginal $p(\mathbf{y})$ calculated over all samples. We report the mean and standard deviation of IS across splits, providing both a point estimate and uncertainty quantification. For binary classification, IS ranges from 1.0 to 2.0.

**Structural similarity index:** To assess the diversity of generated images, we compute pairwise SSIM [2] between all generated samples. For images $\mathbf{x}_i$ and $\mathbf{x}_j$, SSIM is defined as

$$\text{SSIM}(\mathbf{x}_i, \mathbf{x}_j) = \frac{(2\mu_i\mu_j + C_1)(2\sigma_{ij} + C_2)}{(\mu_i^2 + \mu_j^2 + C_1)(\sigma_i^2 + \sigma_j^2 + C_2)}, \tag{S16}$$

where $\mu_i$ and $\mu_j$ are mean intensities, $\sigma_i^2$ and $\sigma_j^2$ are variances, $\sigma_{ij}$ is the covariance, and $C_1$, $C_2$ are two variables to stabilize the division with weak denominator. These constants depend on the dynamic range of the pixel values (typically $L = 255$ for 8-bit images): $C_1 = (K_1L)^2 = 6.50$ and $C_2 = (K_2L)^2 = 58.52$ with $K_1 = 0.01$ and $K_2 = 0.03$. SSIM ranges from $-1$ to 1, with 1 indicating identical images. We report the average pairwise SSIM, estimated by Monte-Carlo sampling. Because evaluating all $N(N-1)/2$ pairs is computationally expensive at our working resolution ($N = 2{,}304$ generated images per test set, giving $\sim 2.65 \times 10^6$ pairs), we draw $M = 1{,}000$ pairs $(i_m, j_m)$ uniformly without replacement from the set $\{(i,j) : 1 \leq i < j \leq N\}$ and compute

$$\overline{\text{SSIM}} = \frac{1}{M}\sum_{m=1}^{M} \text{SSIM}(G(\mathbf{x}_{i_m}, \mathbf{y}_{i_m}), G(\mathbf{x}_{j_m}, \mathbf{y}_{j_m})). \tag{S17}$$

Lower SSIM values indicate higher diversity among generated outputs, while higher values suggest the model produces similar images. This metric complements IS by directly measuring output variation rather than classifier confidence.

**Digit preservation accuracy:** We measure the percentage of generated images correctly classified as their intended digit by the pre-trained classifier:

$$\text{DPA} = \frac{1}{N}\sum_{i=1}^{N} \mathbb{1}[\arg\max(C(G(\mathbf{x}_i, \mathbf{y}_i))) = \mathbf{y}_i], \tag{S18}$$

where $\mathbb{1}$ is an indicator function returning 1 if the condition is true and 0 otherwise. This metric directly quantifies semantic correctness: whether the transformation preserves digit identity.

**Pixel variance:** To further quantify the diversity of generated images, we compute the pixel-wise variance across all generated samples. For a set of $N$ generated images, the pixel variance at position $(h, w)$ is defined as

$$\mathrm{Var}_{h,w} = \frac{1}{N} \sum_{i=1}^{N} \left( G(\mathbf{x}_i, \mathbf{y}_i)_{h,w} - \bar{\mu}_{h,w} \right)^2 , \tag{S19}$$

where $G(\mathbf{x}_i, \mathbf{y}_i)_{h,w}$ denotes the pixel value at position $(h, w)$ of the $i$-th generated image, and $\bar{\mu}_{h,w} = \frac{1}{N} \sum_{i=1}^{N} G(\mathbf{x}_i, \mathbf{y}_i)_{h,w}$ is the mean pixel value at that position. We report the average pixel variance across all spatial locations,

$$\overline{\mathrm{Var}} = \frac{1}{H \times W} \sum_{h=1}^{H} \sum_{w=1}^{W} \mathrm{Var}_{h,w}, \tag{S20}$$

where $H$ and $W$ are the image height and width, respectively. Higher pixel variance indicates greater diversity in the generated outputs, while lower values suggest the model produces similar images across different inputs.

# Supplementary Note 5 Methods

The initial dataset consisted of four grayscale 28×28 standard Helvetica images each of ‘0’ and ‘1’, which differed from each other only in tiny skews and translations of the digits. Computer-Generated Holograms (CGHs) corresponding to each of these eight standard images were prepared with suitable grating parameters and resolution of a HOLOEYE Pluto-2.1 phase-only Spatial Light Modulator (SLM), illuminated by a pulsed laser beam (515 nm wavelength). Each standard image can be expressed as a discrete map $f(p_x, p_y) : [\![1, 28]\!]^2 \to [0, 1]$ where $(p_x, p_y)$ denotes the co-ordinate of pixel $p$ and $[\![1, 28]\!] \equiv [1, 28] \cap \mathbb{Z}$. Hence, $f$ represents the standard image as pixelwise grayscale values. The SLM is a pixel-wise diffractive screen, such that its output intensity profile in image plane can be described convolutionally in terms of some input $f$ describing the target image, and the Fourier transforms involved to produce it. Effectively, however, we can model the SLM’s output spatial profile (of the first-order diffracted light) in its image plane to be of the form

$$I_0(x, y) = \sum_{p_x=1}^{28} \sum_{p_y=1}^{28} f(p_x, p_y) \phi(x - x_p, y - y_p) \equiv \phi * f, \tag{S21}$$

where $(x_p, y_p) := (\Delta X \cdot p_x + x_0, \Delta Y \cdot p_y + y_0)$ denotes the center of the pixel $(p_x, p_y)$ in the image plane. $(x_0, y_0)$ is controlled by the SLM grating parameters, and $\Delta X,\ \Delta Y$ are represented by the SLM resolution. Hence, the SLM image $I_0(x, y)$ is a convolution of the discrete standard image $f(p_x, p_y)$ with some point-spread function $\phi(x, y)$ inherent to the SLM.

Since we are using a pulsed laser, our SLM spatiotemporal output $I(x, y, t) \equiv I_0(x, y)\mathcal{T}_\nu(t)$ has the form $\mathcal{T}_\nu(t) \propto \exp(-\frac{(t-(t_0+\nu T))^2}{2\tau^2})$ where $I_0(x, y) = I(x, y, t = t_0 + \nu T)$ is identified, $T$ represents the duty cycle, $\nu \in \mathbb{N}$, and $t_0 < T$ is arbitrary. In our experiment, Light Conversion CARBIDE laser was used, with a temporal pulse width $\tau \approx 214$ fs.

Due to the stochastic nature of our polariton experiment, each pulsed excitation period must admit some spatial noise (due to jitters in optical setup, etc.), as well as temporal fluctuations in the laser profile $\mathcal{T}_\nu(t)$. In Supplementary Note 6 we show that the polaritonic structural noise leading to GAN performance advantage is crucially due to temporal fluctuations preserved by nonlinearity.

We shall now describe the procedure of our experiment. Firstly, we describe the fabrication process of our microcavity sample hosting the polariton condensate. Secondly, the methodology of preparing the GAN training set (latent prior) is described via three different protocols: **(a)** latent prior is generated by adding digital ii.d. Gaussian random noise to each of the standard $28 \times 28$ grayscale images of '0' and '1', i.e. the set of discrete maps $\{f_i : i \in [\![1, 8]\!]\}$, **(b)** latent prior is generated by adding digital ii.d. Gaussian random noise to the bare SLM projection corresponding to each standard image, i.e. the set $\{\phi * f_i\}$, and **(c)** latent prior is polariton enriched, i.e. the variation across the training set is due to structured noise generated from polariton nonlinearity, when the desired SLM output image is incident on our microcavity sample. More specifically, the latent prior in protocol **(c)** can be described to be generated from the set $\{\mathcal{K}_M * \phi * f_i :\ i \in [\![1, 8]\!],\ M \in [\![1, 480]\!]\}$, where the exact relation between $\mathcal{K}_M$ and the non-linear polariton kernel $\mathcal{K}$ (defined in Eq. S42 of Supplementary Note 6) will be discussed later.

## Supplementary Note 5.1 Microcavity sample preparation

Our planar microcavity sample is composed of a thin polymethyl methacrylate (PMMA) film containing R3B-SMILES planar dipoles, sandwiched between a top silver mirror and a bottom $SiO_2/TiO_2$ distributed Bragg reflector (DBR) on quartz substrate. This system hosts a Tamm plasmon mode within the interfacial region between the metal and DBR, overlapping the SMILES layer.

The SMILES complex is prepared from Rhodamine 3B (R3B) dye, a fluorescent dye commonly used as a gain medium in dye lasers, with an excitonic transition at $\sim 566$ nm. We combined R3B perchlorate ($ClO_4^-$) and cyanostar

macrocycle in a chlorobenzene solvent. The SMILES solution is mixed with PMMA solution, a polymer that controls the thickness of the R3B-SMILES thin film during the spin-coating process. The detailed synthesis procedure of R3B-SMILES film in PMMA is described in [3].

We deposited the SMILES thin film on the DBR substrate by spin-coating the SMILES solution. The film thickness of $(40 \pm 5)$ nm was designed to create a Tamm cavity mode around a wavelength of 600 nm. The detuning between the cavity mode and the exciton transition at $k_{\|} = 0$, controlled by the film thickness, was designed so that the lower polariton at $k_{\|} = 0$ was resonant with the PL emission maximum. This ensured pumping of the lower polariton from the exciton reservoir. From spectroscopic data, we found that the PL emission from polariton condensation is strongly peaked in the range $(596 \pm 1)$ nm throughout the sample surface.

Finally, to fabricate the top silver mirror, a $\sim 100$ nm thin Ag film was slowly evaporated on top of the SMILES film on DBR substrate, via e-beam vapor deposition (EBPVD).

## Supplementary Note 5.2 GAN Training Dataset preparation

As discussed earlier in Supplementary Note 5, our methodology follows the generation of the latent prior via three different protocols **(a)**, **(b)**, and **(c)**. Each protocol's effect on GAN performance is reported in Results section of our main article. Here we describe each of these protocols.

### Supplementary Note 5.2.1 Protocol (a): Latent prior digitally generated

In this protocol, no optical setup is involved: the latent prior is generated entirely digitally from the set of 8 base Helvetica grayscale images $\{f_i : i \in [\![1, 8]\!]\}$ introduced in Supplementary Note 5. Each base image $f_i$ is first upsampled from its native $28 \times 28$ resolution to the target working resolution of $428 \times 428$ using Lanczos interpolation, yielding an 8-bit grayscale image $F_i$ on the intensity range $[0, 255]$. Pixelwise i.i.d. Gaussian noise is then added to produce 480 perturbed realisations per base image

$$\tilde{F}_i^{(M)}(s_x, s_y) = F_i(s_x, s_y) + \sigma\, \eta_{s_x,s_y}^{(M)}, \qquad \eta_{s_x,s_y}^{(M)} \sim \mathcal{N}(0, 1), \tag{S22}$$

where $(s_x, s_y) \in [\![1, 428]\!]^2$ indexes pixel position, $M \in [\![1, 480]\!]$ indexes the realisation, $\sigma = 150$ is the noise standard deviation, and the resulting values are subsequently clipped to the valid intensity range $[0, 255]$. The full latent prior in this protocol is therefore given by the set $\{\tilde{F}_i^{(M)} : i \in [\![1, 8]\!],\ M \in [\![1, 480]\!]\}$,

from which a total of $8 \times 480 = 3{,}840$ points are sampled. This corresponds to the baseline mode described in Supplementary Note 2, and provides the purely computational reference against which the bare-laser protocol (b) and the polariton-enriched protocol (c) are compared in the main text.

### Supplementary Note 5.2.2 Protocol (b): Latent prior digitally generated from SLM alone

Since our work aims to show a computational advantage in replacing computational noise with polariton-enriched optoelectronic noise, the noise used to generate the latent prior in this case is entirely digital. The spatially modulated laser beam (corresponding to each $f_i$) was focused (objective magnification: 20×) normally onto an ordinary reflective mirror, which was placed on the adjustable stage of an Olympus IX73 microscope. The reflected light was thus captured in real space by a Teledyne Princeton CCD camera with an exposure of one second. The microscope focus was adjusted so as to get a clear SLM projection of each standard image $f_i$ on the camera. To protect the CCD camera and avoid image saturation, the peak pulse power ($I_0$) was set low while maintaining a higher pulse repetition rate (i.e. $1/T$) of 120 Hz. The latter was done to minimize analog optoelectronic noise from the CCD (since noise is added digitally in this protocol). This is because a higher pulse rate gives a stronger signal-to-noise ratio for a fixed exposure time.

The camera that captures the well-focused SLM projection produces a discrete $428 \times 428$ image of the continuous projection $\phi * f_i$. The effect of this can be described by defining a set of $428^2$ sampling basis functions $S = \{\psi_{s_x,s_y}(x,y) : s_x, s_y \in [\![1, 428]\!]\}$. Each $428 \times 428$ frame of the bare SLM image $b_i(M)$, with frame index $M$, corresponding to each $f_i$ can be appropriately defined pixelwise as

$$b_i(s_x, s_y, M) = \left( \sum_{\nu=120M}^{120M+119} \int_{\nu T}^{(\nu+1)T} \mathcal{T}_\nu(t)dt \right) \int\int (\phi * f_i)\psi_{s_x,s_y}(x,y)dxdy \tag{S23}$$

$$[\text{using: } I(x,y,t) := \mathcal{T}_\nu(t)I_0(x,y) \equiv \mathcal{T}_\nu \phi * f_i]$$

$$\sim 120 \times \langle (\phi * f_i), \psi_{s_x,s_y} \rangle \tag{S24}$$

such that $\langle \psi_{s_x,s_y}, \psi_{s'_x,s'_y} \rangle = \delta_{s_x,s'_x}\delta_{s_y,s'_y}$. 480 such frames were collected for each $f_i$. The latent prior vector $\mathbf{x}_b$ is prepared from $\{b_i(M)\}$ by digitally adding a different instance of noise vector $\mathbf{n}$ (of size $428 \times 428$) sampled from pixelwise i.i.d. Gaussian random distribution $\mathcal{N}(\mathbf{0}, \mathbf{1})$, to each $b_i(M)$. Hence, the latent prior in this protocol is described by the distribution $\mathbf{x}_b \sim b_i + \epsilon\mathcal{N}(\mathbf{0}, \mathbf{1})$, where $\epsilon$ is some arbitrary small real number, from which a total $480 \times 8 = 3{,}840$ points are sampled.

## Supplementary Note 5.2.3 Protocol (c): Latent prior generated from polariton structured noise

In this protocol, we simply replace the reflective mirror on the microscope stage with a $\sim 1$ cm$^2$ wafer of our microcavity sample pasted on a glass slide. The light from the microcavity, comprising both the red PL emission and the green laser excitation, was filtered through a 550 nm long-pass filter. The red light was thus captured in real space by our CCD camera with 1 s exposure time per frame.

We operated at a low pulse repetition rate of 5 Hz, and the pulse power was set just high enough to observe condensate photoluminescence in the clear shape of the corresponding digit ($f_i$), with added structured noise across frames due to polariton nonlinearity, when properly focused. Hence, each captured frame $M$ can be described by the $428 \times 428$ image $c_i(M)$, corresponding to each $f_i$, which is a composition of PL emission from five pulses as

$$c_i(s_x, s_y, M) = \left\langle \left( \frac{5\gamma}{2\pi D} K_0 * \phi * f_i + \sum_{\nu=5M}^{5M+4} \int_{\nu T}^{(\nu+1)T} (\mathcal{K} \circledast I) dt \right), \psi_{s_x,s_y} \right\rangle \tag{S25}$$

[from Eqs. S34, S38, & S41]

$$= \left\langle \left( \frac{5\gamma}{2\pi D} K_0 * \phi * f_i + \sum_{\nu=5M}^{5M+4} \int_{\nu T}^{(\nu+1)T} \mathcal{K} \circledast (\mathcal{T}_\nu \phi * f_i) dt \right), \psi_{s_x,s_y} \right\rangle \tag{S26}$$

$$= \left\langle \left( \frac{5\gamma}{2\pi D} K_0 + \sum_{\nu=5M}^{5M+4} \int_{\nu T}^{(\nu+1)T} (\mathcal{K} \circledast \mathcal{T}_\delta) dt \right) * (\phi * f_i), \psi_{s_x,s_y} \right\rangle \tag{S27}$$

$$[\mathcal{T}_\delta : \mathcal{T}_\delta(x,y,t) = \mathcal{T}_\nu(t)\delta(x)\delta(y)]$$

$$= \langle (\mathcal{K}_M * \phi * f_i), \psi_{s_x,s_y} \rangle. \tag{S28}$$

Since each frame is a composition of PL emission from only five pulses, the structured variations across the frames is prominent. 480 such frames were captured for each of the eight images in the initial standard dataset. Hence, the training set consisted of $480 \times 8 = 3{,}840$ images.

Due to the low pulse repetition rate in this protocol, the presence of uncorrelated analog optoelectronic noise is significant. However, such unstructured random effects can be assumed to be modeled linearly frame-by-frame, such that the polariton-enriched latent prior can be described by the set of points $\{c_i(M) + \mathcal{A}_M : i \in [\![1,8]\!],\ M \in [\![1,480]\!]\}$, where $\mathcal{A}_M$ denotes a vector of size $428 \times 428$ sampled from uncorrelated optoelectronic noise characteristic of the CCD device.

# Supplementary Note 6 Role of nonlinearity in polaritonic advantage

In this work, we propose a hybrid analog-digital GAN architecture, in which the latent prior of a traditional c-GAN is enriched with structured noise resulting from polariton dynamics. We can understand this in the language of convolutions, namely of Fourier optics, since our microcavity size is much smaller compared to our overall free-space setup, and all optical elements experience near-normal incidence of the collimated laser beam. Typically, Fourier optics is valid for linear systems imaged in the far field, but the effect of nonlinearity on our analysis can be qualitatively understood in a perturbative manner.

Before we move on to the significance of nonlinearity, we can analytically investigate what happens if the polariton cloud response to an excitation at time $t$ and at a point $(x, y)$ in the excitation plane (SLM image plane) is described by a space–time kernel (Green's function) $G(x-x', y-y', t-t')$ in the linear-response approximation, which is reasonable in the pumping regime slightly above threshold. Then the time-dependent polariton density $n(x, y, t)$ (or emitted PL intensity per unit time, up to a proportionality) produced by the optical excitation $I(x, y, t)$ being convoluted by $G$, is given by

$$n(x,y,t) = \int\int\int G(x-x', y-y', t-t')I(x', y', t')dx'dy'dt' \tag{S29}$$

where $G : \mathcal{L}G(\mathbf{s}, \mathbf{s'}) = \delta^3(\mathbf{s}-\mathbf{s'})$, $\mathcal{L}$ being some linear operator governing polariton dynamics, i.e. $\mathcal{L}n(\mathbf{s}) = I(\mathbf{s})$ ($\mathbf{s} \equiv (x, y, t)$ and likewise for $\mathbf{s'}$). For instance, if we adopt a semiclassical diffusion+decay model with source $I(x, y, t)$ such that

$$\partial_t n = D\nabla^2 n - \gamma n + I(x, y, t) \tag{S30}$$

where $\gamma$ is the polariton decay rate and $D$ is some parameter akin to the effective mass governing polariton dynamics, then the corresponding $G(\mathbf{s}, \mathbf{s'})$ can be evaluated as

$$G(r, t-t') = \frac{1}{4\pi(t-t')D}\exp[-\frac{r^2}{4(t-t')D} - \gamma(t-t')], \tag{S31}$$
$$r = \sqrt{(x-x')^2 + (y-y')^2}$$

The time-integrated photoluminescence (PL) measured over the $\nu^{th}$ pulse for an exposure $T$ is

$$P_{out}(x, y, \nu) \propto \gamma \int_{\nu T}^{(\nu+1)T} n(x, y, t)dt$$

$$= \gamma \int\int \left( \int_{\nu T}^{(\nu+1)T} \int G(x-x', y-y', t-t') \mathcal{T}_\nu(t') dt' dt \right) I_0(x', y') dx' dy' \tag{S32}$$

$$= \int\int K_\nu(x-x', y-y') I_0(x', y') dx' dy' \tag{S33}$$

$$\equiv K_\nu * I_0 = K_\nu * (\phi * f) \tag{S34}$$

where $*$ is the convolution operator as expressed previously, except now it is defined on continuous maps. Following through with our strictly linear model, we assume $\mathcal{T}_\nu(t') \sim \frac{1}{\sqrt{2\pi}\tau} e^{-(t'-(t_0+\nu T))^2/2\tau^2} \sim \frac{1}{T}\left[\delta(\frac{t'-t_0}{T}-\nu) + \mathcal{O}_\nu\left((\frac{\tau}{T})^2\right)\frac{\partial^2}{\partial(\frac{t'}{T})^2}\delta(\frac{t'-t_0}{T}-\nu) + ...\right]$ as $\frac{\tau}{T} \ll 1$, such that two distinct pulses are temporally identical only upto $\mathcal{O}(1)$ terms. The $\mathcal{O}_\nu(\tau^2)$ and higher order terms represent pulsewise temporal fluctuations in the laser profile that integrate out to zero over timescale $T$ (stochasticity). Thus we obtain

$$K(r,\nu) \approx \gamma \Bigg[ \int_{\nu T}^{(\nu+1)T} \int G(r, t-t') \delta(t'-t_0-\nu T) dt' dt + \mathcal{O}_\nu(\tau^2) \frac{\partial}{\partial t} G(r, t-t_0-\nu T) \Bigg|_{\nu T}^{(\nu+1)T} + ... \Bigg] \tag{S35}$$

$$= \gamma \Bigg[ \int_{\nu T}^{(\nu+1)T} \frac{1}{4\pi(t-t_0-\nu T)D} e^{-\frac{r^2}{4(t-t_0-\nu T)D} - \gamma(t-t_0-\nu T)} dt + \mathcal{O}_\nu(\tau^2) \partial_s G(r,s) \Bigg|_{-t_0}^{T-t_0} + ... \Bigg] \tag{S36}$$

$$= \gamma \left[ \frac{1}{4\pi D} \int_{-\infty}^{ln\left(\sqrt{\frac{4D}{\gamma}}\frac{T}{r}\right)} e^{-\sqrt{\frac{\gamma}{D}} r \cdot cosh(z)} dz + \mathcal{O}_\nu(\tau^2) \partial_s G(r,s) \Bigg|_0^T + ... \right] \tag{S37}$$

$$\left[ \text{chosen: } t_0 = 0, \;\; z = ln\left( \sqrt{\frac{4D}{\gamma}} \frac{t-t_0-\nu T}{r} \right) \right]$$

$$= \frac{\gamma}{2\pi D} K_0(\sqrt{\frac{\gamma}{D}} r) + 0 + ... \qquad [\text{as} \;\; T \to \infty] \tag{S38}$$

where $K_0$ is the modified Bessel function of the second kind. The main take-away is that for a linear system, identical pulses $I(x,y,t)$ sufficiently spaced in time (i.e. $\tau << T$) will produce identical time-integrated PL of the form Eq. S33, since the convolutional kernel $K(r,\nu)$ is independent of pulse index $\nu$. Importantly, the higher order $\mathcal{O}(\tau^\alpha)$ ($\alpha \in \mathbb{N}$) terms do not contribute pulsewise since $\partial_s^{\alpha-1} G(r,s)|_0^\infty$ terms identically vanish $\left(\partial_s^\alpha \equiv \frac{\partial^\alpha}{\partial s^\alpha}\right)$. The effect of added polariton nonlinearity can be qualitatively understood as follows. Firstly, the

response due to a weak non-linear convolution of an optical impulse can be expressed in terms of the $n^{th}$-order Volterra Kernels ($h_n$) [4],

$$P_{out}(x,y,\nu) \propto \gamma \int_{\nu T}^{(\nu+1)T} n(x,y,t)dt$$
$$= \gamma \left[ \int_{\nu T}^{(\nu+1)T} (G \circledast I)dt + \int_{\nu T}^{(\nu+1)T} ((h_2 \circledast I) \circledast I)dt + ... \right] \quad \text{(S39)}$$

where we define

$$h_n \circledast I =$$
$$\int h_n(x-x_1, x-x_2, .., x-x_n;\ y-y_1, .., y-y_n;\ t-t_1, .., t-t_n) I(x_1,y_1,t_1) dx_1 dy_1 dt_1, \quad \text{(S40)}$$

and we identify $h_1 \equiv G(x-x_1, y-y_1, t-t_1)$ (check Eq. S32), the Green's function from the linear treatment as described. It is clear from this description that while we may approximate $P_{out}(x,y,\nu) = \int_{\nu T}^{(\nu+1)T} (G \circledast I)dt \sim K_0 * I_0$ (from Eqs. S34 & S38) in the strictly linear case, we certainly cannot do so for a general non-linear system when history-dependent kernels $h_n$ become significant, i.e. for a system with transient memory due to non-linear convolutions of $\mathcal{O}(\tau^\alpha)$ contributions to the temporal impulse profile. As a result, $P_{out}(x,y,\nu)$ may depend on $\nu$ even for identical pulses, but only perturbatively different from the strictly linear case, dictated by the polaritonic system's weak $h_n$ for $n \geq 2$. We may express such convolutional fluctuations $\Delta P_{out}(x,y,\nu) \sim P_{out}(x,y,\nu) - \frac{\gamma}{2\pi D} K_0 * I_0$ as

$$\Delta P_{out}(\nu) = \int_{\nu T}^{(\nu+1)T} (\mathcal{K} \circledast I)dt \quad \text{(S41)}$$
$$\text{where}\ \ \mathcal{K} = \gamma \left[ h_2 \circledast I + (h_3 \circledast I) \circledast I + ... \right] \quad \text{(S42)}$$

The set $\{\Delta P_{out}(\nu) : \nu \in \mathbb{N}\}$, whose structure is governed by $\mathcal{K}$, is interpreted as structured noise resulting from nonlinear polariton dynamics. The lengthscale upto which such dynamics are relevant, i.e. the structural noise feature size, can be estimated as $\sqrt{D/\gamma}$ (check Eq. S38).

## Supplementary Note 6.1 Insignificance of spatial noise

Spatial fluctuations in the SLM projection may arise due to spatial jitters in the optical setup, electromechanical vibrations, footsteps, etc. Here we demonstrate via experimental data backed by a simple random walk model that such contributions to the overall optoelectronic noise are negligible in both the bare laser and the polaritonic setup.

We may model the SLM output in the presence of spatial noise as: $I_0(x,y,t) = \sum_{p_x=1}^{28} \sum_{p_y=1}^{28} f(p_x,p_y) \phi(x - x_p + \epsilon_x(t), y - y_p + \epsilon_y(t))$. Here

$(\epsilon_x(t), \epsilon_y(t))$ represents the time-dependent spatial fluctuations. It is assumed that each pulse excitation event admits spatial noise $\epsilon_{x/y}$ randomly chosen pixelwise from Uniform$(-\epsilon, +\epsilon)$, where $\epsilon$ is some maximal fluctuation.

Let each captured frame (indexed by $M$) be a composition of $N \sim 1/T$ pulses, where $T$ is the duty cycle (our exposure time $= 1$ $s$). Hence,

$$\text{Frame}(M) = \left(\sum_{\nu=MN}^{(M+1)N-1} \int_{\nu T}^{(\nu+1)T} \mathcal{T}_\nu(t)dt\right)(\phi * f)$$
$$+ \left(\sum_{\nu=MN}^{(M+1)N-1} \int_{\nu T}^{(\nu+1)T} \mathcal{T}_\nu(t)\Big(\epsilon_x(t), \epsilon_y(t)\Big)dt\right) \cdot (\nabla\phi * f) + ... \tag{S43}$$

using, $I(x, y, t) := \mathcal{T}_\nu(t)I_0(x, y)$, and Taylor expanding $\phi(x-x_p+\epsilon_x, y-y_p+\epsilon_y)$ around $(x - x_p, y - y_p)$. We shall use $\mathcal{T}_\nu(t) := \delta(t - t_0 - \nu T)$ for simplicity. It follows that,

$$\text{Frame}(M) = N \times (\phi * f)$$
$$+ \sum_{\nu=MN}^{(M+1)N-1} \Big(\epsilon_x(t_0 + \nu T), \epsilon_y(t_0 + \nu T)\Big) \cdot (\nabla\phi * f) + ... \tag{S44}$$

Let $S_{X,M} = \sum_{\nu=0}^{N-1} \xi_{X,M}(\nu)$, where $\xi_{X,M}(\nu) \equiv \epsilon_x\Big(t_0 + (\nu + MN)T\Big)$. Likewise, we define $S_{Y,M}$. $\xi_{X,M}$ can be thought of as a 1D random walk variable, where the random walk's step size is uniformly (and randomly) chosen from the real interval $(-\epsilon, +\epsilon)$ for each instance of $\nu$. Hence, $S_{X,M}$ can be thought of as the sum of $N$ random walks, i.e. a random variable denoting the displacement from origin after $N$ random, independent 1D steps. Likewise for $S_{Y,M}$. The probability distribution function over the variable $S_{X,M}$ can be derived as

$$f_{S_{X,M}}(s) = \frac{1}{2\pi}\int_{-\infty}^{\infty}\left(\frac{\sin(k\epsilon)}{k\epsilon}\right)^N e^{-iks}dk \tag{S45}$$

from which we can calculate the variance, $Var(s) = \mathbb{E}(s^2) - \mathbb{E}(s)^2 \sim \mathcal{O}(N\epsilon^2)$, where $\mathbb{E}(.) = \int(.)f_{S_{X,M}}(s)ds$. Hence, larger the number of steps $N$, larger will be the range of possible net displacement(s) after $N$ random steps. For small $N$, the net displacement would be sharply peaked around zero. Our captured frame (SLM image) is of the form,

$$\text{Frame}(M) = N \times (\phi * f) + \Big(\mathcal{O}_M(\sqrt{N}\epsilon), \mathcal{O}_M(\sqrt{N}\epsilon)\Big) \cdot (\nabla\phi * f)$$
$$+... \tag{S46}$$

where $\mathcal{O}_M(\sqrt{N}\epsilon) := (-N\epsilon, N\epsilon)\ \forall\ M$. That is, each frame admits pixelwise (i.e., in both $x-$ and $y-$ directions) a spatial noise of size at most $N\epsilon$, where $N$ is the number of pulses integrated per frame, such that the uncorrelated spatial variations across the frames is quantified pixelwise by $\sqrt{N}\epsilon$.

The bare laser protocol **(b)** (in Supplementary Note 5.2.2) and the polariton-enriched protocol **(c)** operated at $N = 120$ and $N = 5$, respectively. Hence, protocol **(c)** should be less affected by such spatial variations than protocol **(b)**. Moreover, as discussed in Fig. S1 below, there is negligible variation across frames obtained via protocol **(b)**, as well as in polaritonic PL data obtained at a larger $N$ ($N = 50$). It is therefore safe to ignore the role of spatial noise throughout the analysis presented in our text.

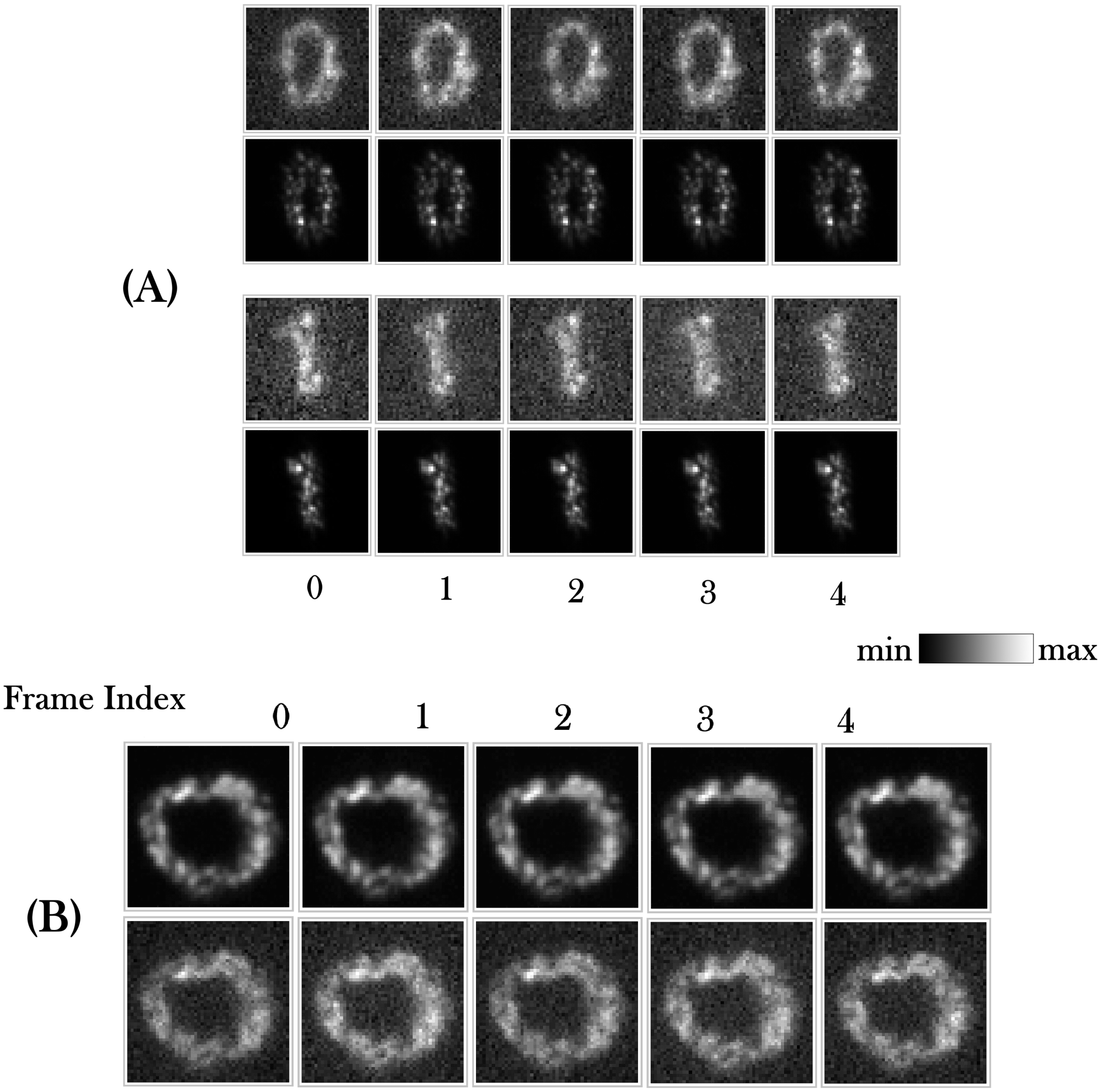


**Supplementary Fig. S1** Comparison between frame-by-frame realisations of **(A)** a standard Helvetica digit ('0' or '1'), as utilised in our main correspondence to generate bare laser and polaritonic latent priors; and of **(B)** an MNIST digit '0' transformed via two different polaritonic protocols (unrelated to our main results). Second and fourth rows of **(A)** represent latent prior images from the bare laser protocol as discussed, whereas the polaritonic PL training data are presented in the first and third rows. In addition to background optoelectronic random white noise due to low signal-to-noise ratio (SNR) in polaritonic PL patterns, there are also structured textural variations in the digit itself, which are absent in the bare laser images with a better SNR. Likewise, polariton-transformed PL patterns each composed of 5 pulsed excitations (**bottom row, (B)**) show familiar structural differences across different realisations of the '0'; whereas compositions of 50 pulses (**top row, (B)**) show enhanced SNR but lose textural variability, despite having a larger number of pulsed excitation cycles ($N$), admitting more spatial fluctuations hypothetically. This highlights the insignificance not only of spatial fluctuations in any case (small or large $N$), but also that of non-linear effects when the time-integrated PL is composed of a larger $N$.

# Supplementary References

[1] Salimans, T., Goodfellow, I., Zaremba, W., Cheung, V., Radford, A., Chen, X.: Improved techniques for training gans. Advances in neural information processing systems **29** (2016)

[2] Wang, Z., Bovik, A.C., Sheikh, H.R., Simoncelli, E.P.: Image quality assessment: from error visibility to structural similarity. IEEE transactions on image processing **13**(4), 600–612 (2004)

[3] Deshmukh, P., Satapathy, S., Michail, E., Olsson, A.H., Bushati, R., Yadav, R.K., Khatoniar, M., Chen, J., John, G., Laursen, B.W., *et al.*: Plug-and-play molecular approach for room temperature polariton condensation. Acs Photonics **11**(2), 348–355 (2024)

[4] Boyd, S., Chua, L.: Fading memory and the problem of approximating nonlinear operators with volterra series. IEEE Transactions on circuits and systems **32**(11), 1150–1161 (1985)


\end{document}